\documentclass[preprint,12pt]{elsarticle}

\usepackage{amsmath}
\usepackage{amssymb}
\usepackage{amsthm}
\usepackage{algorithm}
\usepackage{algpseudocode}
\usepackage{subfig}
\usepackage{url}

\def\N{{\mathbb{N}}}
\def\IMF{{\textrm{IMF}}}
\newtheorem{definition}{Definition}

\makeatletter
\def\@author#1{\g@addto@macro\elsauthors{\normalsize%
    \def\baselinestretch{1}%
    \upshape\authorsep#1\unskip\textsuperscript{%
      \ifx\@fnmark\@empty\else\unskip\sep\@fnmark\let\sep=,\fi
      \ifx\@corref\@empty\else\unskip\sep\@corref\let\sep=,\fi
      }%
    \def\authorsep{\unskip,\space}%
    \global\let\@fnmark\@empty
    \global\let\@corref\@empty  
    \global\let\sep\@empty}%
    \@eadauthor={#1}
}
\makeatother

\begin{document}

\begin{frontmatter}

\title{On the Instantaneous Phase and Frequency Estimation of a Non-Stationary Multicomponent Signal. The JADE Algorithm}

\author[1]{Jayanth Mouli}
\ead{jmouli@gatech.edu}

\author[1]{David V. Anderson}
\ead{anderson@gatech.edu}

\author[2]{Antonio Cicone\corref{cor1}}
\ead{antonio.cicone@univaq.it}
\cortext[cor1]{Corresponding author}

\address[1]{School of Electrical and Computer Engineering, Georgia Institute of Technology, Atlanta, GA, USA}
\address[2]{DISIM, University of L'Aquila, L’Aquila, Italy; Istituto di Astrofisica e Planetologia Spaziali, INAF, Rome, Italy; Istituto Nazionale di Geofisica e Vulcanologia, Rome, Italy.}

\begin{abstract}
Many real-life signals, such as gravitational wave measurements, biomedical signals, or geophysical data, are strongly non-stationary but can be decomposed into mono-component signals that contain only one active frequency over time. This is made possible thanks to decomposition methods developed in recent years that can handle non-stationary signals. The problem now is how to compute, in an accurate and stable way, the instantaneous frequency, phase, and amplitude of such mono-component signals. Numerous approaches have been developed so far, but they can be unstable in the presence of noise and struggle to capture quick and intrawave changes in frequency. In this work, we present an alternative approach, called the JADE method, which is based on the Dynamic Time Warping algorithm and which we combine with the FIF algorithm to handle and study multicomponent non-stationary signals. We test the robustness of JADE  to noise and run comparisons with classical methods used for instantaneous frequency, phase, and amplitude estimation.
\end{abstract}

\end{frontmatter}

\section{Introduction}
Real-life signals are non-stationary in general. They are generated by complex and nonlinear systems. In recent years, many innovative algorithms have been developed that are able to decompose real-life signals into simple oscillatory components. We can think, for instance, of Empirical Mode Decomposition \cite{huang1998empirical}, Ensemble Empirical Mode Decomposition \cite{wu2009ensemble}, Fast Iterative Filtering \cite{barbarino2021stabilization, cicone2020iterative, lin2009iterative}, sparse time-frequency representation \cite{hou2011adaptive}, Geometric Mode Decomposition \cite{yu2018geometric}, Variational Mode Decomposition \cite{dragomiretskiy2013variational},  and Empirical Wavelet Transform \cite{gilles2013empirical}, just to mention a few.

Those decomposition methods based on iterations (Empirical Mode Decomposition, Ensemble empirical mode decomposition, and Fast and Resampled Iterative Filtering) have the advantage that they do not require any {\em a priori} assumptions on the number of components to be extracted and do not require the selection of a basis to be used in the decomposition, like in other methods which are based on optimization. Nevertheless, Empirical Mode Decomposition, Ensemble empirical mode decomposition, and derived algorithms, even though widely used in many applications \cite{stallone2020new}, still lack a mathematical foundation \cite{chui2024spline, huang2005introduction}. For this reason, we opt to use Fast Iterative Filtering (FIF) to decompose signals.

For the sake of completeness, we briefly recall here the FIF algorithm and its main features. The goal of FIF is to decompose a given signal $s$ into Intrinsic Mode Functions (IMFs).
We start by defining a filter as a function $w$ that is a nonnegative, even, and compactly supported function with area equal to one and support on $[-L,\ L]$. $L$ is called the filter length. Given a signal $s$, the main step in the algorithm is given by
\begin{equation}\label{eq:FIF}
  s_{m+1} = s_m - s_m \ast w, \qquad m\in\N,
\end{equation}
where $s_1=s$, and $s \ast w$ represent the convolution of the signal with the filter $w$. The filter length $L$ is calculated based on the signal local extrema relative positions \cite{cicone2021numerical}. Iteration \eqref{eq:FIF} is repeated until a stopping criterion is fulfilled, {\em e.g.}, $\left\|s_m-s_{m+1}\right\|/\left\|s_m\right\|\leq \delta$, for a prefixed $\delta>0$ \cite{lin2009iterative}. Once this criterion is fulfilled, $s_m$ is declared to be a simple oscillatory and non-stationary mono-component, also known as an Intrinsic Mode Function (IMF). An IMF is a function that fulfills two conditions: the number of minima and maxima differ by no more than plus or minus one; the envelope connecting the maxima and the one connecting the minima are symmetric with respect to the horizontal line \cite{huang1998empirical}. The second and subsequent IMFs can be computed using the approach applied to the remainder signal $s-\sum_{k=1}^m\IMF_k$, where $\IMF_k$ represents each of the $m$ IMFs already computed.

In \cite{cicone2021numerical}, it was proven that if the filter $w$ is chosen as a convolution of an even filter function $\widetilde{w}$ with itself, then the convergence of the algorithm can be guaranteed a priori. Furthermore, it is possible to accelerate the computations by working in the frequency domain via the Fast Fourier Transform (FFT). In the frequency domain, convolution becomes a simple product. The algorithmic complexity of FIF method therefore is $O(m n\log(n))$ where $n$ is the length of the signal under study, and $m$ is the number of mono-components extracted. Finally, we recall that the fact that we choose the filter length $L$ based on the signal itself implies that FIF and all the derived methods are non-stationary. In fact, given two signals $p$ and $q$, in general, $\textrm{IMFs}(p+q)\neq \textrm{IMFs}(p)+\textrm{IMFs}(q)$.

Once a signal has been decomposed into IMFs, there is a need to study their frequency content to characterize them. Many papers have been published in the past introducing the concept of instantaneous frequency and reviewing the methods developed for its computation \cite{boashash2015time, chen2002time, loughlin1997comments, soto2023concept, stankovic2014instantaneous, wang1998performance, wu2020current}.

Among the approaches developed so far, we recall here the estimation of the instantaneous frequency of a signal based on the time-frequency representations \cite{cohen1995time, stankovic2014time}, and its variant, which is based on nonparametric estimation based on the maxima position of the time-frequency representation \cite{boashash1992estimating, boashash1992estimating2}. Other approaches are based on the intersection of the confidence intervals \cite{stankovic1998algorithm} and the modified intersection of the confidence intervals \cite{djurovic2004modification} algorithms,  amplitude demodulation based upon the reassignment method \cite{czarnecki2016instantaneous}, Wigner-Ville distribution \cite{rao1990estimation}, cross Wigner-Ville distribution \cite{boashash1993use}, directionally smoothed polynomial Wigner–Ville distribution \cite{barkat2001instantaneous,shui2007instantaneous}, robust Wigner distribution \cite{djurovic2001robust}, spectrogram with varying window length \cite{katkovnik2000instantaneous}, Hilbert Transform \cite{huang1998empirical} and Normalized Hilbert Transform \cite{huang2009instantaneous}, peeling, i.e. the maxima position of the strongest component \cite{djurovic2004estimation}, the generalized zero-crossing (GZC) \cite{huang2009instantaneous}, segmentation processes based on mathematical morphology \cite{khan2012instantaneous, rankine2007if}, Synchrosqueezing transform
\cite{wu2013instantaneous}, IMFogram \cite{barbe2022time, cicone2023IMFogram}, Sinusoidal Fitting Decomposition \cite{nie2024sinusoidal}, Reassignment Operators and Linear Chirp Points Detection \cite{colominas2024instantaneous}. The interested reader can find more details in \cite{wu2020current}.

Even though many different methods have been proposed in the literature, they can be unstable to noise and, more importantly, cannot capture intrawave modulations in the frequency, {\em i.e.} variation that takes place even inside each period of the oscillatory components. All these methods are designed to capture up to interwave modulations in frequency, {\em i.e.} changes in frequency that take place from one period to the next one. However, non-stationary signals are, in general, intrawave modulated \cite{huang2009instantaneous}. As an example, we can consider, for instance, a biomedical signal like the electrocardiogram of a patient \cite{martis2014current}, or an astrophysical signal like a gravitational wave produced by the collision of two black holes in deep space \cite{miller2019new}, or a geophysical signal as the Earth's magnetic field measured through a satellite orbiting around the globe \cite{loto2019goes}. All these signals present rapid changes in their frequency content. To be properly analyzed they require an algorithm able to capture the instantaneous frequency contained in them. In this work, we propose an innovative approach that is based on the so-called Dynamic Time Warping algorithm, and we show its robustness against noise compared to traditional methods available in the literature.

The rest of this work is organized as follows: in Section \ref{sec:methods} we review a few methods for the computation of a monocomponent signal's instantaneous frequency and then propose our approach. In Section \ref{sec:Num_Ex_Art} results relative to a few artificial examples are presented, including the test of the robustness of the proposed method and previously published techniques to noise. Section \ref{sec:real_life} concerns applying the proposed method to a few real-life applications. In the last section, we derive conclusions and highlight future directions of research.

\section{Methods}\label{sec:methods}
\subsection{Instantaneous Frequency Estimation Methods}

Non-stationary signals are characterized by their instantaneous frequency (IF), which is defined as the local frequency of the signal at a specific time. We describe several methods to compute the IF of a monocomponent signal in this section. IF is commonly computed through the analytic signal produced by the Hilbert transform (HT). For an input signal $x(t)$, this approach constructs the complex-valued analytic signal $Z(t) = x(t) + j\hat{x}(t)$, where
\begin{equation}
    \hat{x}(t) =\frac{1}{\pi}P.V.\int_{-\infty }^{+\infty}\frac{x(\tau)}{t-\tau}d\tau
\end{equation}
is the Hilbert Transform of $x(t)$, and the phase $\phi(t)$ is the angle of $Z(t)$ \cite{huang1998empirical}. The IF of the signal, $\omega(t)$, is then obtained from:
\begin{equation}\label{eq:IF}
    \omega(t) = \frac{1}{2\pi}\frac{\mathrm{d} \phi(t)}{\mathrm{d} t}
\end{equation}
Huang {\em et al.}~\cite{huang2009instantaneous} determined that several conditions, including those proposed by Bedrosian \cite{bedrosian1963quadrature} and Nuttall \cite{nuttall1966quadrature}, need to be satisfied to obtain an accurate IF estimate using the HT analytic signal approach. To satisfy these conditions, Huang {\em et al.}~\cite{huang2009instantaneous} introduced the Normalized Hilbert Transform (NHT) approach, in which an empirical method separates amplitude modulation (AM) from frequency modulation (FM). In this method, a cubic spline curve $e_1(t)$ is fit to the envelope of the signal, and this curve is used to normalize the input $x(t)$:
\begin{equation}
    y_1(t) = \frac{x(t)}{e_1(t)},
\end{equation}
where $y_1(t)$ is the normalized signal. This normalization scheme is applied iteratively until all signal values are less than or equal to unity. After the $n$th iteration, the normalized signal
\begin{equation}
    y_n(t) = \frac{y_{n-1}(t)}{e_n(t)},
\end{equation}
It is only frequency modulated, with the AM component removed. The IF of the normalized signal $y_n(t)$ is then computed using the HT analytic signal approach.

Another IF estimation method proposed by Huang {\em et al.}~\cite{huang2009instantaneous} is Direct Quadrature (DQ), which also uses the previously described normalization scheme. Here, assuming the original signal $x(t) = A(t)y_n(t)$, and $y_n(t) = \cos(\phi(t))$, the quadrature of $y_n(t)$ is computed as $\sqrt{1-y_n(t)^2}$. The phase is obtained as:
\begin{equation}\label{eq:DQ}
    \phi(t) = \arctan\frac{y_n(t)}{\sqrt{1-y_n(t)^2}}
\end{equation}
The IF is then computed from the derivative of Eq. \ref{eq:DQ}. The advantage of the DQ approach is that it bypasses the HT integral and solely computes the phase from differentiation.

\subsection{Dynamic Time Warping}

Dynamic Time Warping (DTW) is a widely used algorithm to compare time series data by performing a temporal alignment. DTW is capable of determining the optimal alignment despite differing lengths and time shifts of input signals. Suppose there are two time series $\textbf{X}=[x_1, ..., x_n]$ and $\textbf{Y}=[y_1, ..., y_m]$ of lengths $n$ and $m$ respectively. The two series can be aligned to form a cost matrix $\textbf{D}\in \mathbb{R}^{n\times m} $. The cost matrix \textbf{D} is initialized under two constraints:
\begin{itemize}
  \item $\textbf{D}_{i, 0} = \infty $ for $i\in[1,n)$ and $\textbf{D}_{0, j} = \infty$ for $j\in[1,m)$
  \item $\textbf{D}_{0, 0} = 0$
\end{itemize}
The local distance measure $d(\cdot , \cdot  )$ used in this study is the Euclidean distance. The matrix \textbf{D} is then populated according to the formula:
\begin{equation}
    \textbf{D}_{i,j}=d(\textbf{X}_i, \textbf{Y}_j) + \min
    \begin{Bmatrix}
    \textbf{D}_{i-1,j-1} \\ \textbf{D}_{i-1,j}
     \\\textbf{D}_{i,j-1}
    \end{Bmatrix}
\end{equation}
A warping path $\textbf{P}$ is a set containing the indices of the aligned elements: $\textbf{P} = [(i_1, j_1),..., (i_L, j_L)] $. $\textbf{P}$ is defined as a path through the cost matrix \textbf{D} that satisfies the following conditions:
\begin{itemize}
  \item $\textbf{P}_1 = (i_0, j_0) = (0,0)$
  \item $\textbf{P}_L = (i_L, j_L) = (n-1,m-1)$
  \item $i_{k-1}\le i_{k}\le(i_{k-1}+1)$ for $k\in(1,L)$
  \item $j_{k-1}\le j_{k}\le(j_{k-1}+1)$ for $k\in(1,L)$
\end{itemize}
The cost $C_P$ of a given path $\textbf{P} = [(i_1, j_1),..., (i_L, j_L)] $ is defined as:
\begin{equation}\label{eq:cost}
    C_P = \sum_{k=1}^{L}\textbf{D}_{i_k,j_k}
\end{equation}
With the set of all possible paths \textbf{P} defined as $\mathbb{P}_{n,m}$, the optimal path $\boldsymbol\Omega$ is:

\begin{equation}
    \boldsymbol\Omega = \underset{\textbf{P}}{\arg\min} \left\{ C_P |\textbf{P}\in \mathbb{P}_{n,m}\right\}
\end{equation}
This path $\boldsymbol\Omega$ can be determined by tracing back through the cost matrix from $\textbf{D}_{n-1, m-1}$ to $\textbf{D}_{0,0}$. The cost $C_{\boldsymbol\Omega} = DTW(\textbf{X}, \textbf{Y})$ of $\boldsymbol\Omega$ is often used as a metric of similarity between the inputs \textbf{X} and \textbf{Y}.

\begin{figure}[ht!] 
\centering
\includegraphics[width=0.8\textwidth]{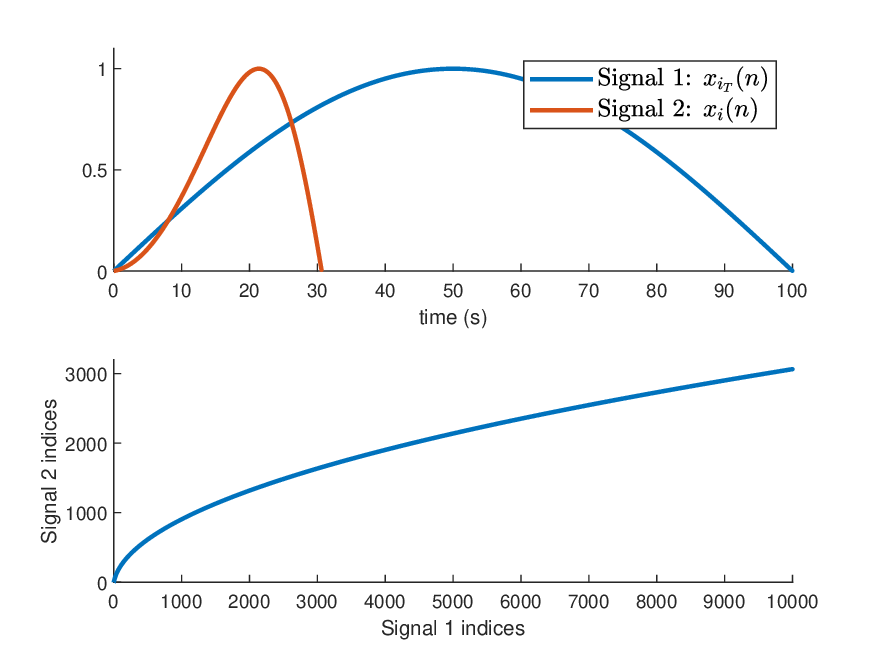}
\caption{Example of DTW alignment between two signals. The original signals (top) are aligned according to the warping path (bottom).}
\label{DTW}
\end{figure}

\subsection{JADE algorithm}
In this study, we use the alignment path $\boldsymbol\Omega$ computed using DTW to estimate the instantaneous phase and frequency of an input signal. We call this newly developed approach the JADE algorithm, from the initials of the author's names and the word estimator. Consider a sampled signal $x(n)=\cos(\phi(n))$, where $T_s$ is the sampling period. Assume $x(n)$ is time-warped using DTW to align with a discrete-time template signal $x_T(k) = \cos(\omega_TkT_s)$, where $\omega_T$ is the known template frequency. Then $x(n)$ can be expressed as $\cos(\omega_T\psi(n))$ where $\psi(n) = \frac{\phi(n)}{\omega_T}$. After the DTW alignment, the optimal warping path $\boldsymbol\Omega(k)$ approximates the following relation:
\begin{equation}
    x(\boldsymbol\Omega(k)) = \cos(\omega_T\psi(\boldsymbol\Omega(k))) \approx \cos(\omega_TkT_s).
\end{equation}

Therefore, $\psi(\boldsymbol\Omega(k)) = T_sk$, and:
\begin{equation}\label{eq:sampling}
    \boldsymbol\Omega^{-1}(n) = \frac{1}{T_s}\psi(n)=\frac{1}{\omega_TT_s}\phi(n)
\end{equation}

By time-warping $x(n)$ into a known template signal using DTW, the phase $\phi(n)$ can be recovered using the inverse of the warping path $\boldsymbol\Omega(k)$. We can now introduce the following two definitions

\begin{definition}[Instantaneous phase $\phi(n)$ of a monocomponent signal $x(n)$]\label{def-iPhase}
\begin{equation}\label{eq:iPhase}
   \phi(n) = \omega_T T_s  \boldsymbol\Omega^{-1}(n)
\end{equation}
where the warping path $\boldsymbol\Omega(k)$ is obtained by time-warping the signal $x(n)$ into a known template signal using DTW.
\end{definition}

\begin{definition}[Instantaneous frequency $\omega(n)$ of a monocomponent signal $x(n)$]\label{def-iFreq}
\begin{equation}\label{eq:iFreq}
   \omega(n) = \frac{1}{2\pi}\frac{\mathrm{d} \phi(n)}{\mathrm{d} n}
\end{equation}
where $\phi(n)$ is the instantaneous phase of the monocomponent signal $x(n)$ given in Definition \ref{def-iPhase}.
\end{definition}

The discrete-time signal model we use to represent a zero-mean oscillatory signal $x(n)$ is $x(n) = \cos(\phi(n)) + w(n)$, where $w(n)$ is Gaussian noise, and the function $\phi(n)$ is to be estimated. The noise can cause several problems when finding $\phi(n)$. It is likely, in fact, that the mono-component signals may contain some noise or secondary frequencies whose energy is lower than the one of the main frequency. This is true both if we make the assumption that the signal under investigation is monocomponent in the first place, or if it has been obtained as a monocomponent IMF via the FIF algorithm.

In our investigations, we observed that when $x(n)$ is non-monotonic, the noise can have an impact on $\phi(n)$ calculation obtained as in \eqref{eq:iPhase}, particularly around the inflection points.  Accordingly, we may select to break the signal into monotonic segments for further analysis.  Alternatively, we can break the signal into segments using the zero crossing locations.  From now on, we opt to use this second approach. However, the same results can be obtained by splitting the signal into monotonic segments.

We denote the  $k$ zero-crossings of $x(n)$ as $\textbf{z} = \left\{ z_1,...,z_k \right\}$. Because the noise $w(n)$ can interfere with zero-crossing detection, a smoothed version of $x(n)$, $x_s(n)$, is created by applying a moving average window with a fixed window length. A heuristic estimates the window length that attenuates 25\% of the energy of the signal $x(n)$. The zero-crossings \textbf{z} are then obtained from the resulting signal. The phase of the signal is estimated starting with the first zero-crossing of the signal, as the phase is known at that location.

Next, $x(n)$ is split into sections $\textbf{x}_k = \left\{ x_1(n),...,x_k(n) \right\}$ at zero-crossing indices $\textbf{z} = \left\{ z_1,...,z_k \right\}$. Each section $x_i(n)\in\textbf{x}_k$ is defined by:
\begin{equation}
    x_i(n) =
    \begin{cases}
        x(n+z_i) & \text{if } 0\le n\le (z_{i+1}-z_i)\\
        0 & \text{if }  n>(z_{i+1}-z_i)
    \end{cases}
\end{equation}
Each section $x_i(n)\in\textbf{x}_k$ resembles a half-period sinusoid, which we denote the template sinusoid $x_{i_T}(n)$. The template sinusoid has amplitude $A_{i_T}$ and frequency $\omega_{i_T}$. The frequency of each template, $\omega_{i_T} = \frac{\pi}{z_{i+1} - z_i}$, is an approximation of the frequency of the corresponding section.

Then, the template sinusoid $x_{i_T}(n)$ can be expressed as:
\begin{equation}\label{eq:tempForm}
    x_{i_T}(n) =
    \begin{cases}
        \pm A_{i_T}\sin(\omega_{i_T}n) & \text{if } 0\le n\le (z_{i+1}-z_i)\\
        0 & \text{if }  n>(z_{i+1}-z_i)
    \end{cases}
\end{equation}
where $x_{i_T}(n)$ matches the sign of $x_i(n)$. If we split the signal at local extrema to create monotonic segments, the template sinusoid will have the form $A_{i_T}\sin(\omega_{i_T}n  \pm \frac{\pi}{2})$.

For zero-mean artificial signals, it is usually sufficient to set the template amplitude $A_{i_T}$ to the maximum or minimum value of the current segment. However, for low-SNR signals or cases in which secondary frequencies may be present, we choose the $A_{i_T}$ that corresponds to the lowest-cost DTW alignment between the template and the current segment:
\begin{equation}\label{eq:tempAmp}
     A_{i_T} = \underset{A}{\arg\min} \left\{ C_{\boldsymbol\Omega} = DTW(x_{i_T}(n), x_i(n))\right\},
\end{equation}
where $C_{\boldsymbol\Omega}$ is the DTW alignment cost, and $x_{i_T}(n) = A  \sin(\omega_{i_T}n)$.

Because $x_i(n)$ and $x_{i_T}(n)$ are similar, DTW can find a low-cost alignment between these two signals, which can help create a smoother warping path for phase retrieval. Equation \eqref{eq:sampling} requires the inverse of the warping path $\boldsymbol\Omega_i^{-1}(k)$, which is determined by inverting the axes of the warping path of $DTW(x_{i_T}(n),x_i(n))$. The section phase estimate is then computed as $\hat{\phi_i}(k) = \omega_{i_T}T_s\boldsymbol\Omega_i^{-1}(k)$ from Eq.(\ref{eq:sampling}), and finally the estimates $\hat{\phi_i}(k)$ are concatenated to obtain the entire phase estimate $\hat{\phi}(k)$ for the input signal $x(n)$.

Because the DTW warping path and its inverse can contain vertical segments, $\hat{\phi}(k)$ is not differentiable. We seek a differentiable version of $\hat{\phi}(k)$, $\hat{\phi}_s(k)$,  in order to compute an IF estimate. To do this, we fit interpolating cubic splines \cite{wolberg1988cubic} of the form $\hat{\phi}_{s_i}(k) = \sum_{n=0}^{3}\alpha    _nk^n$ to $\hat{\phi}(k)$ at partition points $k_i$, which are taken to be the zero-crossings \textbf{z} by default. The coefficients of the cubic polynomial are determined such that the boundary and derivative conditions are satisfied: $\hat{\phi}_{s_i}(0) = \hat{\phi}(k_i)$, $\hat{\phi}_{s_i}(k_{i+1} - k_i) = \hat{\phi}(k_{i+1})$, $\hat{\phi}'_{s_i}(k_{i+1} - k_i) = \hat{\phi}'_{s_{i+1}}(0)$, and $\hat{\phi}''_{s_i}(k_{i+1} - k_i) = \hat{\phi}''_{s_{i+1}}(0)$. The curves $\hat{\phi}_{s_i}(k)$ are concatenated to form a piecewise polynomial phase estimate $\hat{\phi}_s(k)$. Because the derivatives of $\hat{\phi}_s(k)$ are continuous, the IF estimate can be computed as $\hat{\omega}_s(k) = \frac{1}{2\pi}\frac{\mathrm{d} \hat{\phi_s}(k)}{\mathrm{d} k}$. The overall approach is summarized in Algorithm \ref{alg:cap}.

\begin{algorithm}[ht]
\caption{JADE($x(n)$, $T_s$, $n_i$).}\label{alg:cap}
\textbf{Inputs:} Target signal $x(n)$, sampling period $T_s$, cubic spline partition points $n_i$.\\
\textbf{Output:} Total phase estimates $\hat{\phi}_s(n)$, IF estimate $\hat{\omega}_s(n)$
    \begin{algorithmic}[1]
    \State \textbf{z} $\leftarrow$ Zero-crossings of $x(n)$
    \State $\textbf{x}_k = \left\{ x_1(n),...,x_k(n) \right\} \leftarrow$ $x(n)$ split at crossings \textbf{z}
    \State $\hat{\phi}(n)\leftarrow$ $\left\{  \right\}$
    \State $k\leftarrow$ length of \textbf{z}
    \For {$i=1,2,\ldots,k$}
        \State $\omega_{i_T} \leftarrow$ template frequency  $\frac{\pi}{z_{i+1} - z_i}$
        \State $A_{i_T} \leftarrow$ template amplitude from Eq.(\ref{eq:tempAmp})
        \State $x_{i_T} \leftarrow A_{i_T} \sin(\omega_{i_T} n)$
        \State $\boldsymbol\Omega_i\leftarrow$ warping path of $DTW(x_{i_T}(n),x_i(n))$
        \State $\hat{\phi}_i(n)\leftarrow$ phase computed as $\omega_{i_T}T_s\boldsymbol\Omega_i^{-1}(k)$
        \State $\hat{\phi}(n)\leftarrow$ $\left\{ \hat{\phi}(n), \hat{\phi}_i(n)\right\}$
    \EndFor
    \State $\hat{\phi}_s(n) \leftarrow$ cubic spline with partition points \textbf{z}
    \State $\hat{\omega}_s(n) \leftarrow \frac{1}{2\pi}\frac{\mathrm{d} \hat{\phi_s}(n)}{\mathrm{d} n}$
    \State \Return $\hat{\phi_s}(n), \hat{\omega_s}(n)$
    \end{algorithmic}
\end{algorithm}

\section{Analysis of Artificial Signals}\label{sec:Num_Ex_Art}

To demonstrate the potentiality of the proposed JADE method, we first consider its performance on four artificial signals and compare it with alternative methods previously proposed in the literature\footnote{JADE code and all examples reported in this work are available for download at \url{www.cicone.com}}. The first two examples are mono-component signals containing noise, therefore we can apply JADE directly and test its robustness to noise. The other two artificial examples contain two non-stationary frequency components, hence we first apply FIF to decompose them into IMFs and then apply JADE.

\subsection{Artificial example 1}

\begin{figure}[ht!] 
\centering
\includegraphics[width=0.8\textwidth]{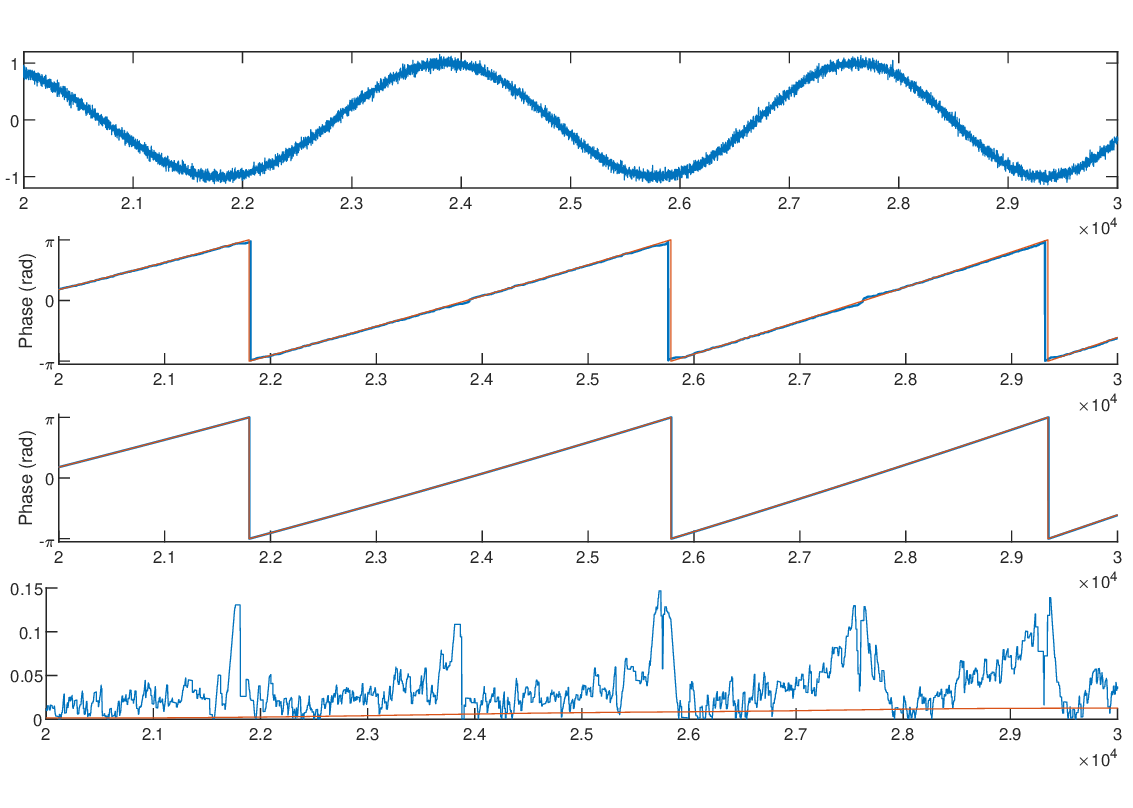}
\caption{Example 1. Phase estimation of a section of the quadratic-phase signal with added Gaussian noise (top). The JADE phase estimate (blue) and ground truth phase (red) are shown without spline fitting (second panel) and with spline fitting (third panel). The bottom panel shows the error between prediction and ground truth phase with spline fitting (red) and without (blue).}
\label{LinearChirp}
\end{figure}

The simplest case to consider is the linear chirp signal with quadratic phase and amplitude 1: $x(n) = \cos(\alpha n^{2}+\beta n) + \gamma \xi_n$, where $\xi_n$ are Gaussian random numbers with zero mean and standard deviation 1, and $\gamma$ is a scaling factor. In Fig.~\ref{LinearChirp}, phase estimation is shown on a quadratic phase signal with $\gamma$ = 0.05, $\alpha$ = $\frac{\sqrt{5}}{1000}$ and $\beta$ = $\frac{\sqrt{2}}{300}$. The Signal to Noise Ratio (SNR), computed as SNR = $20  \cdot log(\left \| signal \right \|_2/\left \| noise \right \|_2)$, of this signal is 23.18 dB. In Fig.~\ref{LinearChirp}, we show the result of the JADE phase estimate with and without spline fitting, in the second and third panels respectively. In both cases, it can be seen that the phase estimated using JADE matches the ground truth phase function of the signal well. However, it is observed that spline fitting improves the accuracy of the phase estimate. This is seen in the fourth panel, where the norm-1 error between the phase estimate and ground truth is plotted for both cases. The error without spline fitting is higher than with spline fitting on average.

Different values of SNR were tested to determine the algorithm's resilience to noise. We define the relative error $\epsilon$ to be:
\begin{equation}\label{eq:error}
    \epsilon = \frac{\left \| \phi-\hat{\phi} \right \|_2}{\left \| \phi \right \|_2},
\end{equation}
where the ground truth phase $\phi(n) = \alpha n^{2}+\beta n$, and $\hat{\phi}(n)$ is the JADE phase estimate. When testing performance over a range of values of SNR, ground truth zero crossings are used. This ensures that the performance of the JADE algorithm is evaluated independently of zero-crossing detection. The assumption of ground truth zero crossings is close to reality up to considerable levels of noise. Table \ref{table_err_SNR} shows measured values of $\epsilon$ for different SNR.

\begin{table}[ht!]
\begin{center}
\begin{tabular}{||c c c c c c c c||}
 \multicolumn{8}{|c|}{\textbf{SNR (dB)}} \\
 \hline
 &25.55 & 13.62 & 9.19 & 4.11 & -1.45 & -6.36 & -10.86 \\ [0.5ex]
 \hline\hline
 \textbf{$\epsilon$}&1.5e-4 & 5.5e-4 & 8.9e-4 & 1.2e-3 & 1.9e-3 & 2.7e-3 & 4e-3  \\
 \hline
\end{tabular}
\end{center}
\caption{Example 1. Relative error in phase estimate (\textbf{$\epsilon$}) versus noise level}
\label{table_err_SNR}
\end{table}

\subsection{Artificial example 2}

\begin{figure}[ht!] 
\centering
\includegraphics[width=0.8\textwidth]{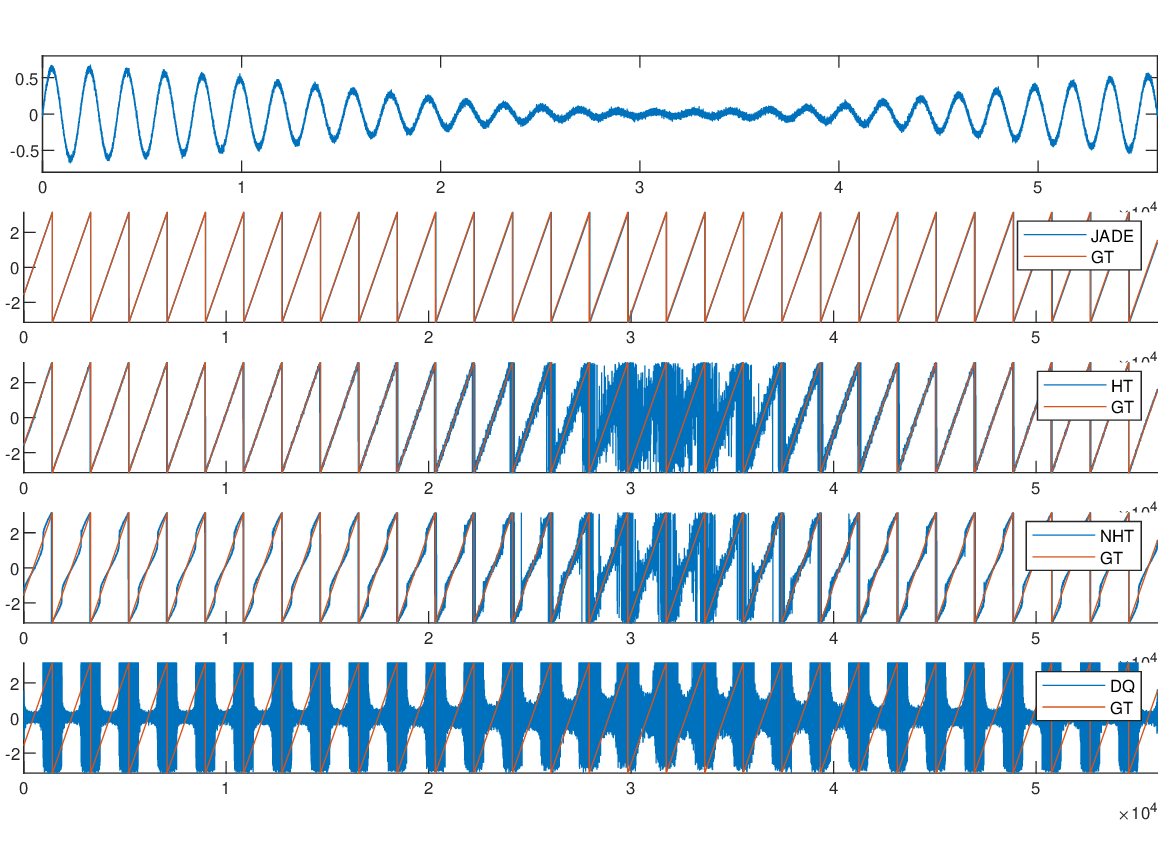}
\caption{Example 2. Analysis of the amplitude and phase modulated signal in Eq. \ref{eq:nonStationary} (top). The JADE phase estimate (blue) and ground truth phase (red) are compared in the second panel. The Hilbert Transform (third panel), Normalized Hilbert Transform (fourth panel), and Direct Quadrature (bottom panel) methods are also compared to ground truth.}
\label{nonStationaryZCHilb}
\end{figure}

The JADE method is next evaluated on an amplitude and frequency modulated artificial signal:
\begin{equation}\label{eq:nonStationary}
    x(n) = [A(n) \cdot \cos(\phi(n))] + \gamma \xi_n,
\end{equation}
where $A(n) = A_1cos(\omega_1n)$, $\phi(n) = n + A_2cos(\omega_2n)$, and $ \gamma \xi_n$ are Gaussian random numbers, $\mathcal{N}(0,1)$, scaled by $\gamma$. This signal is amplitude and phase-modulated, as seen in the top panel of Fig.~\ref{nonStationaryZCHilb}. It is difficult to obtain an expression for the ground truth phase of $x(n)$ from Equation \eqref{eq:nonStationary}. To obtain the ground truth phase, we assume that multi-component signals can be considered single-component if the amplitude and frequency components of the signal can be separated. Adapting Definition 3.1 from Daubechies {\em et al.}~\cite{daubechies2011}, we assume the components of a function $f(n) = A(n) \cdot \cos(\phi(n))$ can be separated if:
\begin{equation}\label{eq:monoconditions}
    \left | A'(n) \right |, \left | \phi''(n) \right |\leq \epsilon \left | \phi'(n) \right |, \forall n\in\mathbb{Z}
\end{equation}
up to accuracy $\epsilon$. For $x(n)$ in \eqref{eq:nonStationary}, choosing $\epsilon$ = 0.01, $A_1$ = 0.3, $\omega_1$ = 1/35, $A_2$ = 1, $\omega_2$ = 1/100, and $\gamma$ = 0.05, the conditions in \eqref{eq:monoconditions} are satisfied. Therefore, for this signal, we assume that the ground truth phase is $\phi(n) = n + \cos(n/100)$.
In the second panel of Fig.~\ref{nonStationaryZCHilb} the JADE phase estimate is shown in blue, along with the ground truth phase $\phi(n)$ in red, wrapped to the range $[-\pi,\pi]$. We also compare the result to the phase estimate based on the Hilbert Transform (third panel), NHT (fourth panel), and DQ (bottom panel) methods. It can be seen that as the amplitude of the signal decreases, the HT, NHT, and DQ phase estimates become erratic, while the JADE phase estimate accurately represents the ground truth phase. This could be because the normalization scheme common to NHT and DQ methods involves spline-fitting, and the addition of noise can impact this method, so the form of the normalized signal is significantly altered. The relative error $\epsilon$ as defined in Eq. \eqref{eq:error} between the JADE estimate and the ground truth phase is 0.055.

\subsection{Artificial example 3}

\begin{figure}[ht]
\centering
    \subfloat[]{\includegraphics[width=0.45\linewidth]{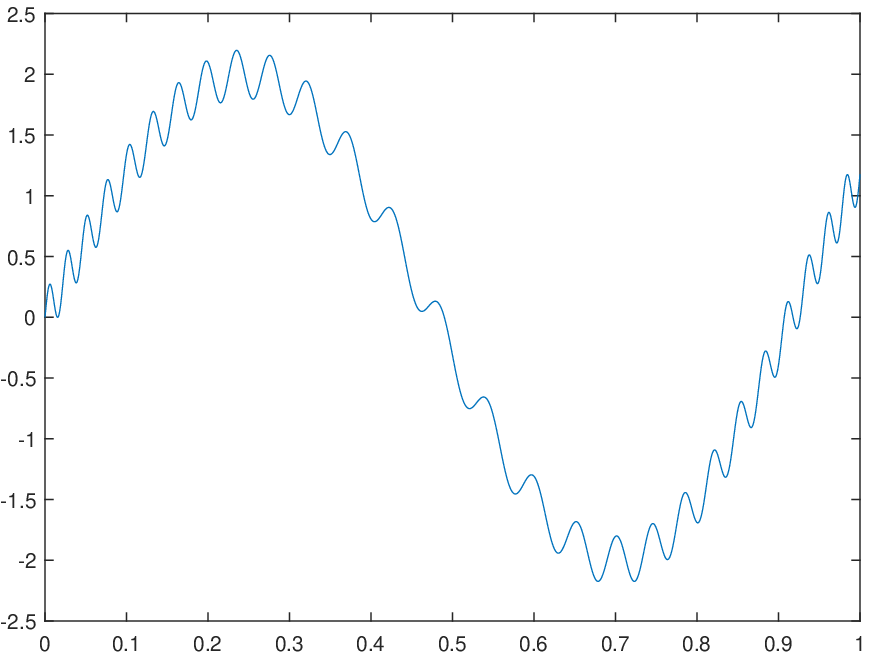}}
\hfil
    \subfloat[]{\includegraphics[width=0.45\linewidth]{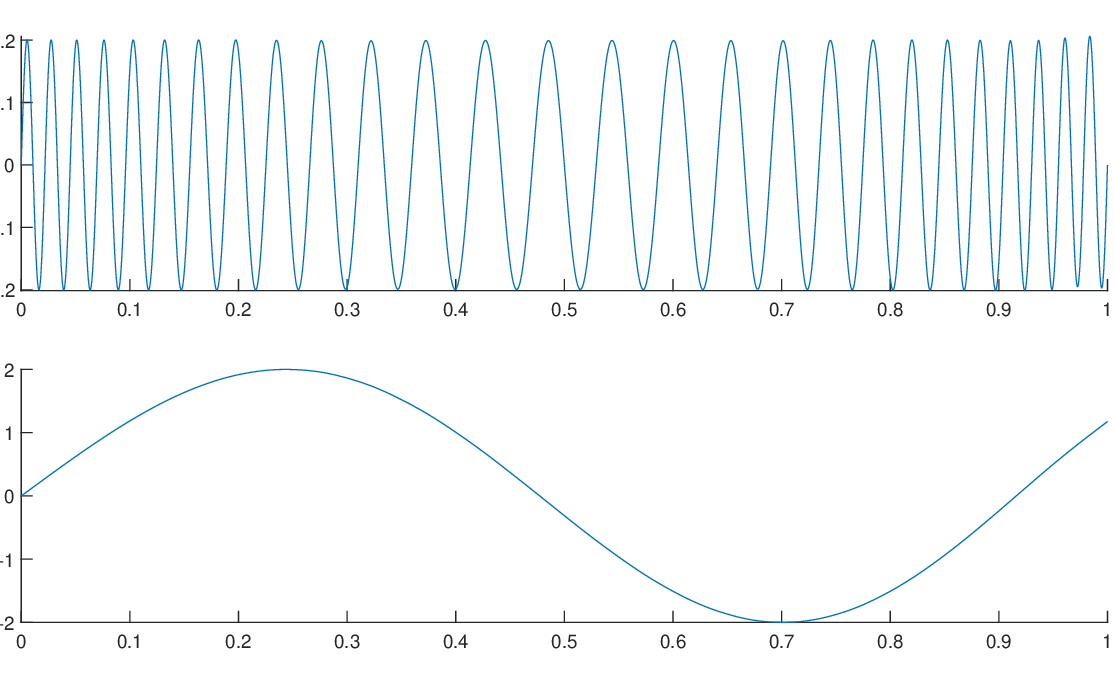}}

    \subfloat[]{\includegraphics[width=0.45\linewidth]{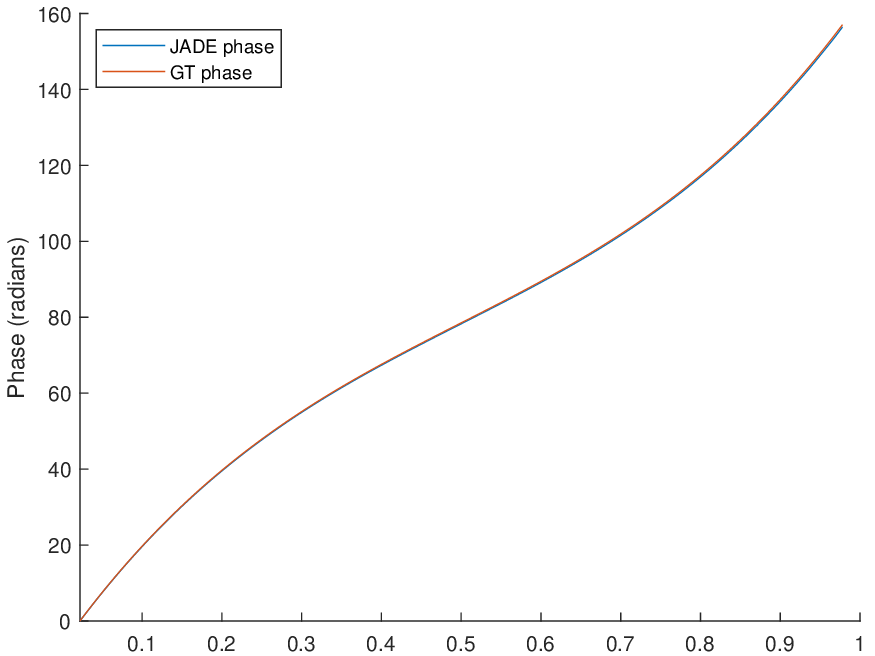}}
\hfil
    \subfloat[]{\includegraphics[width=0.45\linewidth]{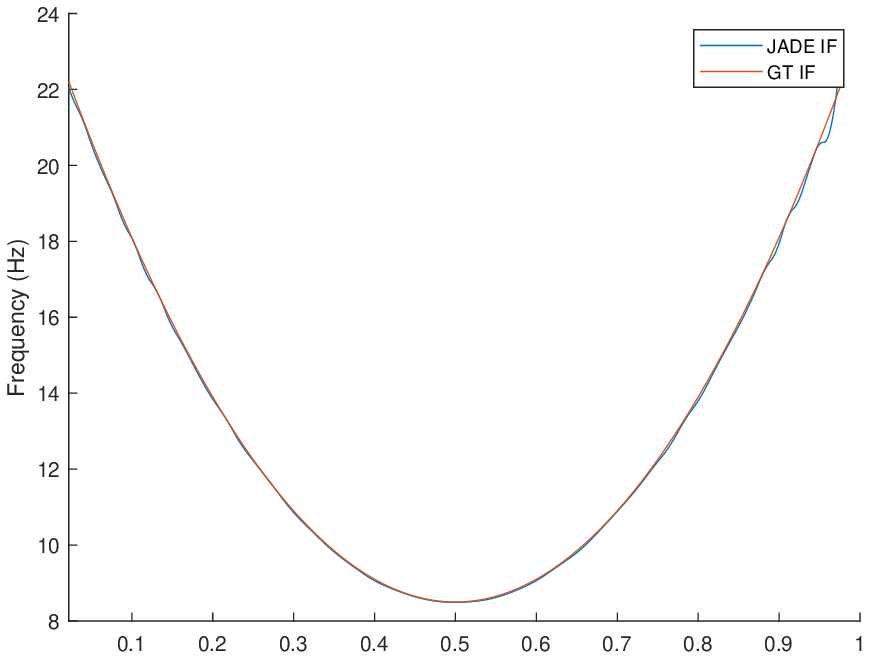}}

    \subfloat[]{\includegraphics[width=0.45\linewidth]{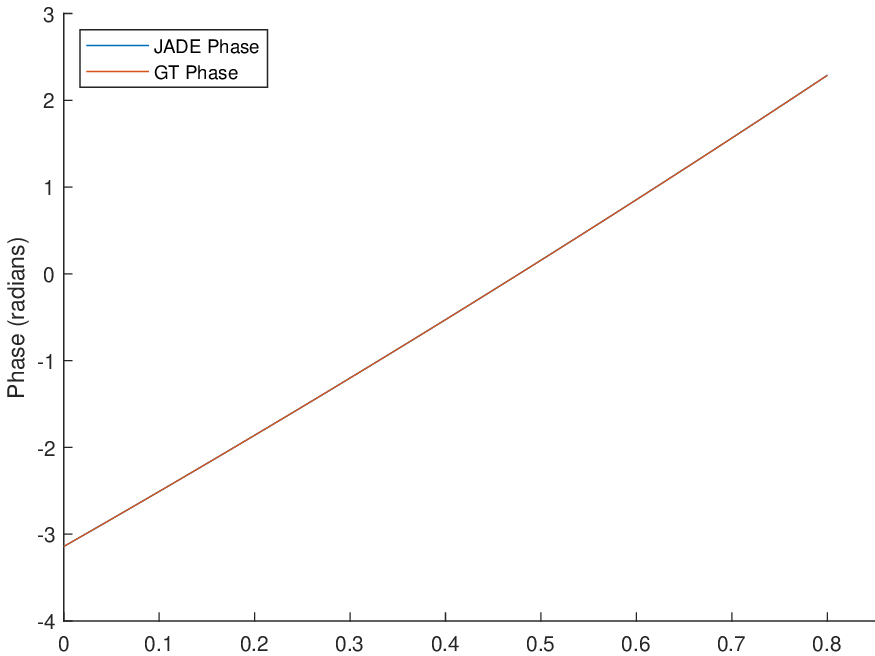}}
\hfil
    \subfloat[]{\includegraphics[width=0.45\linewidth]{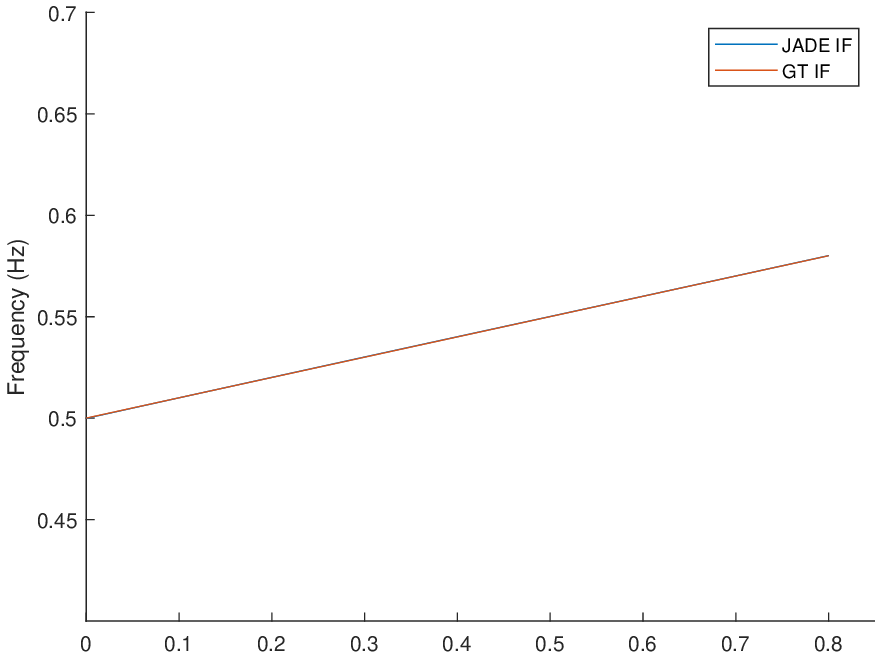}}
\caption{Example 3. (a) A signal with two non-stationary components. (b) IMFs obtained from the FIF method. (c) JADE phase estimate (blue) versus ground truth phase (red) for IMF 1. (d) JADE IF estimate (blue) versus ground truth IF (red) for IMF 1. (e) JADE phase estimate (blue) versus ground truth phase (red) for IMF 2. (f) JADE IF estimate (blue) versus ground truth IF (red) for IMF 2.  }
\label{fig:artificial-multicomponent}
\end{figure}

In this example, we consider a signal containing two non-stationary components whose equations are as follows:
\begin{equation}\label{eq:ex3}
\begin{split}
x(n) & = [0.2 \cdot \sin(\phi_1(n))],\\
y(n) & = [ 2 \cdot \sin(\phi_2(n))], \\
s(n) & = x(n)+y(n),
\end{split}
\end{equation}
where $\phi_1(n) = 2 \pi (40 n^3-60 n^2 + 47 n +$, $\phi_2(n) = 2 \pi (0.1 n^2 + n)$. The derived signal $s$ is depicted in panel (a) of Fig.~\ref{fig:artificial-multicomponent}, whereas its FIF decomposition is reported in panel (b) of the same figure. In the subsequent panels, the IF and phase reconstruction of the two IMFs are presented and compared with the ground truth values. From these results, we see that the combination of the FIF and JADE algorithms allows the reconstruction of the signal IFs and phases.

\clearpage

\subsection{Artificial example 4}

As a last artificial example, we consider the solution of the undamped Duffing equation
\begin{equation}
    \ddot{x}+\alpha x + \beta x^3 = \gamma \cos(\omega t),
\end{equation}
which can be rewritten as a spring-mass system with a forcing function
\begin{equation}\label{eq:duffing}
    \ddot{x}+x k(x)  = \gamma \cos(\omega t)
\end{equation}
where $k(x)=\alpha + \beta x^2$ is a nonlinear spring coefficient which varies with the position $x$ of the mass. Equation \eqref{eq:duffing} can also be interpreted as representing a pendulum with a forcing function and a pendulum length that varies with the angle of the swing. No matter how we interpret  the equation physically, the solution to \eqref{eq:duffing} will be oscillatory with an intrawave modulation in frequency within each period. Continuous changes in the spring constant or length of the pendulum cause constant changes in instantaneous frequency.

\begin{figure}[ht] 
\centering
\includegraphics[width=0.8\textwidth]{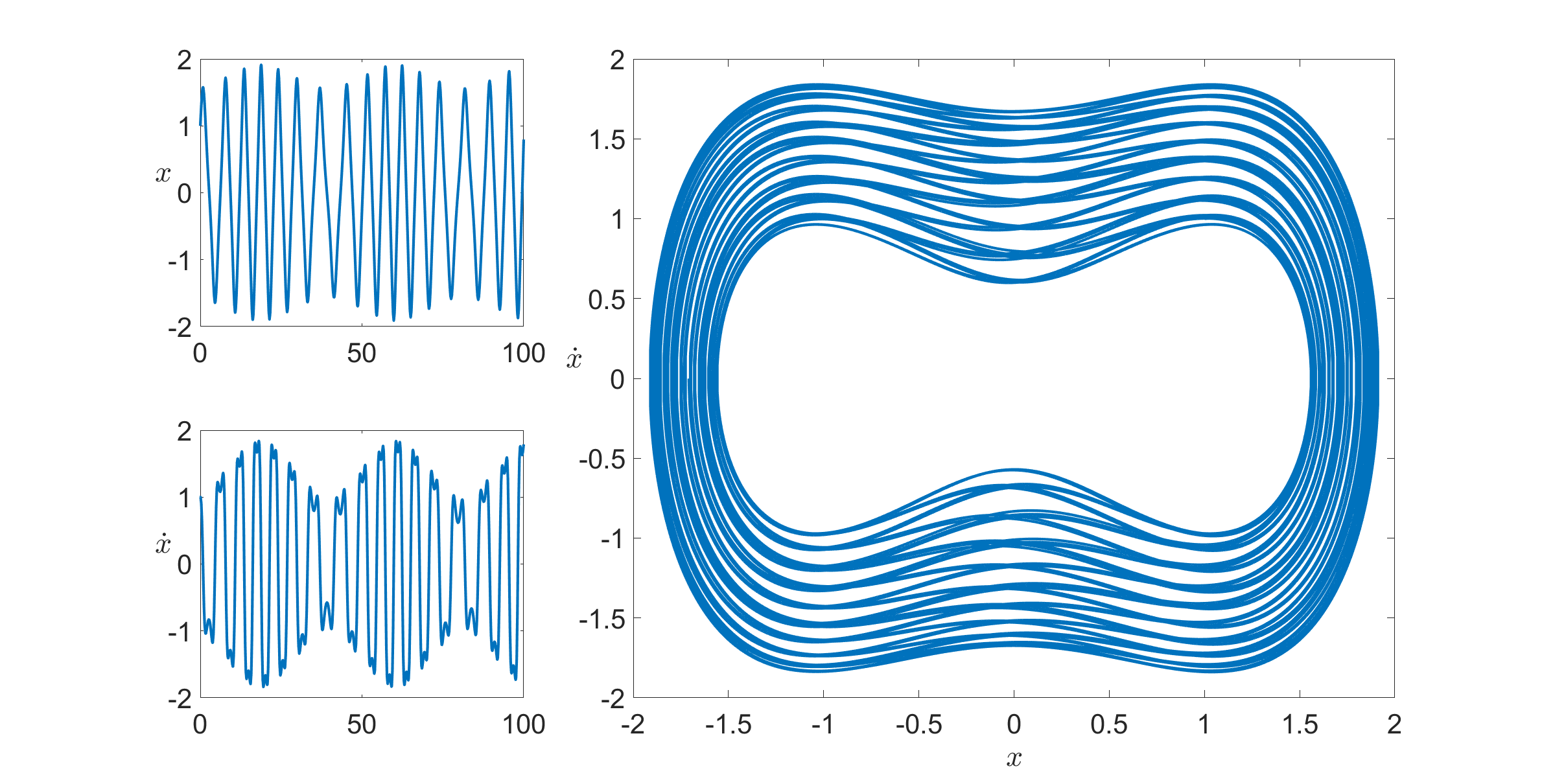}
\caption{Example 4. Solution of the undamped Duffing equation for  $\alpha=-1$, $\beta = 1$, $\gamma = 0.1$, and $\omega= 1$.}
\label{fig:Duffing_sol}
\end{figure}

\begin{figure}[ht] 
\centering
\includegraphics[width=0.8\textwidth]{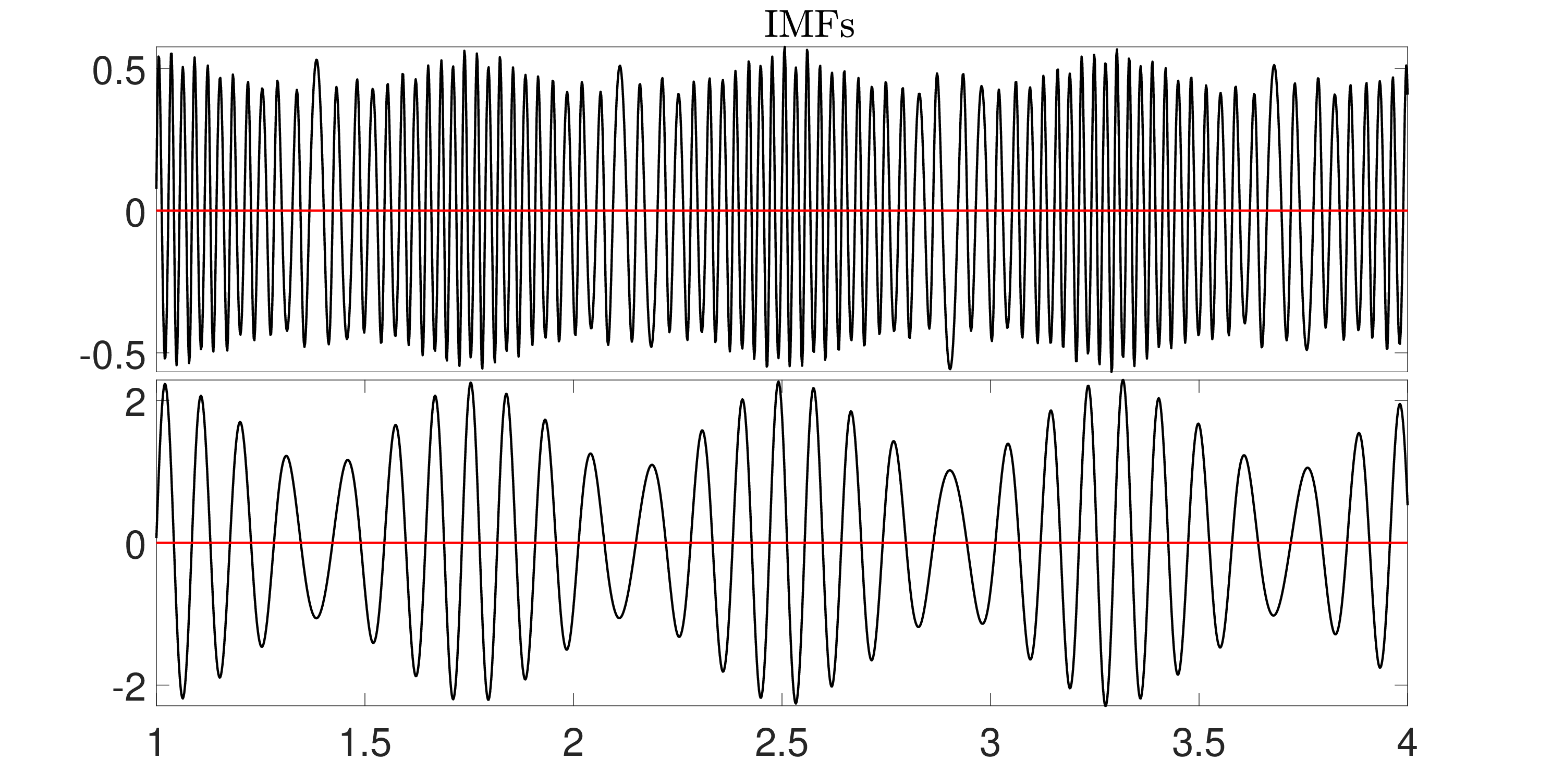}
\caption{Example 4. FIF decomposition into IMFs of $\dot x$ solution of the first 4 minutes of the undamped Duffing equation plotted in Fig.~\ref{fig:Duffing_sol}.}
\label{fig:Duffing_IMFs}
\end{figure}

If we set $\alpha=-1,\, \beta = 1, \, \gamma = 0.1 , \, \omega= 1$, we obtain, through standard numerical integration, the solutions $x$ and $\dot x$ plotted in Fig.~\ref{fig:Duffing_sol}. By considering $\dot x$ as the signal we want to analyze, we can decompose it into simple oscillatory components, IMFs, \cite{huang1998empirical}. We can obtain this decomposition, for instance, using the Iterative Filtering algorithm \cite{cicone2021numerical}. Each IMF is a mono-frequency signal. The decomposition is shown in Fig.~\ref{fig:Duffing_IMFs}.

\begin{figure}[ht] 
\centering
\includegraphics[width=0.8\textwidth]{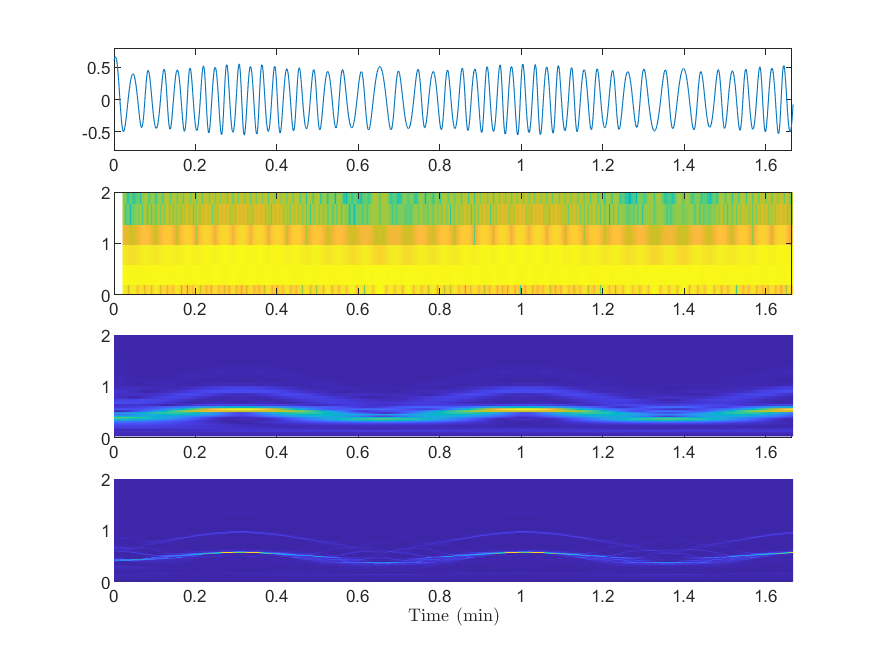}
\caption{Example 4. First IMF of the $\dot x$ solution of the undamped Duffing equation \eqref{eq:duffing} (top panel) and IF estimates from STFT (second), CWT (third), and SST (bottom) in Hertz.}
\label{fig:Duffing_IF_1}
\end{figure}

If we focus our attention on the first 100 seconds of the first IMF extracted from the signal $\dot x$, and we apply the Short Time Fourier Transform (STFT), also known as Spectrogram, the Continuous Wavelet Transform (CWT) \cite{flandrin1998time}, and the Synchrosqueezing Transform (SST) \cite{daubechies2011}, we obtain the time-frequency plots reported in Fig.~\ref{fig:Duffing_IF_1}.

\begin{figure}[ht] 
\centering
\includegraphics[width=0.8\textwidth]{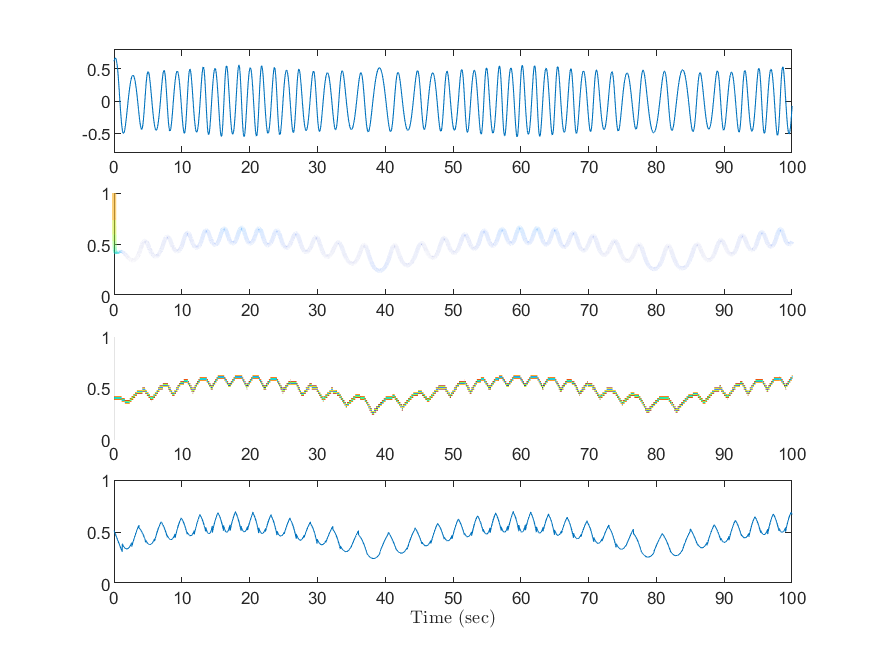}
\caption{Example 4. First IMF of the $\dot x$ solution of the undamped Duffing equation \eqref{eq:duffing} (top panel) and IF estimates from HT (second), IMFogram (third), and JADE (bottom) in Hertz.}
\label{fig:Duffing_IF_2}
\end{figure}

Similarly, on the same signal, we can apply HT \cite{huang2009instantaneous}, IMFogram \cite{barbe2022time, cicone2023IMFogram}, and the proposed JADE algorithm, ref.~Fig.~\ref{fig:Duffing_IF_2}.

From a direct comparison of Fig.~\ref{fig:Duffing_IF_1} and Fig.~\ref{fig:Duffing_IF_2} we can observe that classical methods like STFT, CWT, and the advanced SST have a hard time following the quick local variability in the frequency of this particular signal. These methods rely on bases whose elements, taken individually, cannot represent such a signal. This is the reason why we see more than one frequency curve appearing in all time-frequency planes in Fig.~\ref{fig:Duffing_IF_1}.

Whereas, using HT, IMFogram, and JADE algorithm we end up having a single frequency curve over time, Fig.~\ref{fig:Duffing_IF_2}. These last three methods present similar performance. However, we know from previous tests that among the three, only JADE can produce instantaneous curves, up to numerical precision, and that only JADE is stable to possible noise.

Lastly, we show the full analysis of the multicomponent Duffing equation solution using JADE and FIF. We decompose the original signal into 2 IMFs and obtain phase predictions from JADE for both IMFs. Because the ground truth phase of each IMF is unknown, the performance of the JADE phase estimate cannot be directly evaluated. Instead, we perform a reconstruction of the original signal using the JADE phase estimate. We aggregate the sinusoidal template amplitudes into an amplitude function $\hat{A}(n)$, and also construct a function $\hat{\mu}(n)$ that contains the mean values of each section which were removed to perform the phase estimation. Then, to reconstruct the signal from the JADE phase estimate $\hat{\phi}(n)$, we evaluate:
\begin{equation}\label{eq:reconstruction}
    \widehat{IMF} = [\hat{A}(n) \cdot \cos(\hat{\phi}(n))] + \hat{\mu}(n)
\end{equation}

\begin{figure}[ht] 
\centering
\includegraphics[width=0.8\textwidth]{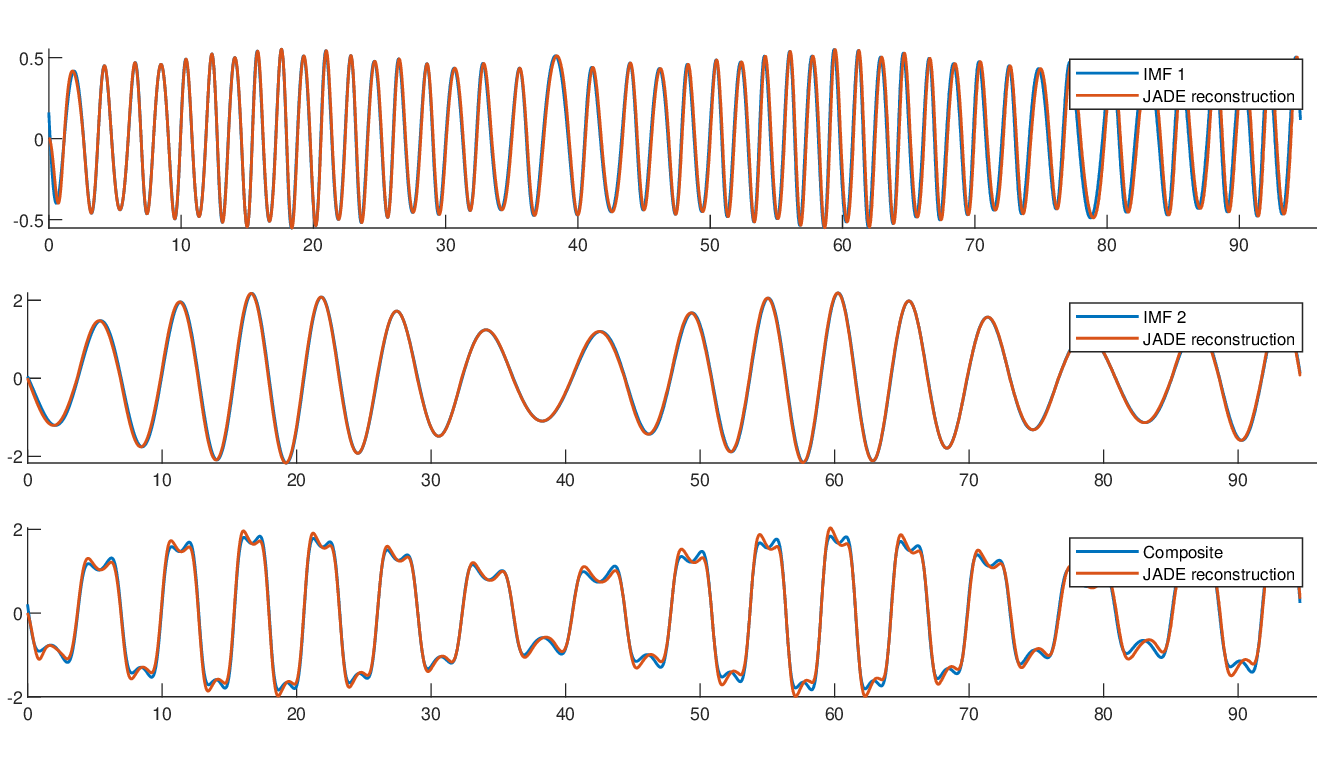}
\caption{Example 4. Reconstructions from the JADE phase estimate using Equation \eqref{eq:reconstruction} for IMF 1 (top) and IMF 2 (middle), in red, with the original IMFs shown in blue. The bottom panel shows the composite sum of JADE reconstructions for each IMF in red, compared with the Duffing equation solution in blue.}
\label{fig:Duffing_JADE_FIF}
\end{figure}

The reconstructions $\widehat{IMF_1}$ and $\widehat{IMF_2}$ of the first two IMFs is shown in Fig.~\ref{fig:Duffing_JADE_FIF}. In the bottom plot of Fig.~\ref{fig:Duffing_JADE_FIF}, we also show the comparison between $(\widehat{IMF_1} + \widehat{IMF_2})$ and the original Duffing solution.

\clearpage

\section{Analysis of real-life examples}\label{sec:real_life}

To demonstrate the abilities of the JADE algorithm, we apply it to analyze electrocardiogram (ECG) signals, gravitational wave (GW) signals, and geomagnetic response waveforms.

\subsection{Example 5: Bird chirping}

\begin{figure}[ht] 
\centering
\includegraphics[width=2.0in]{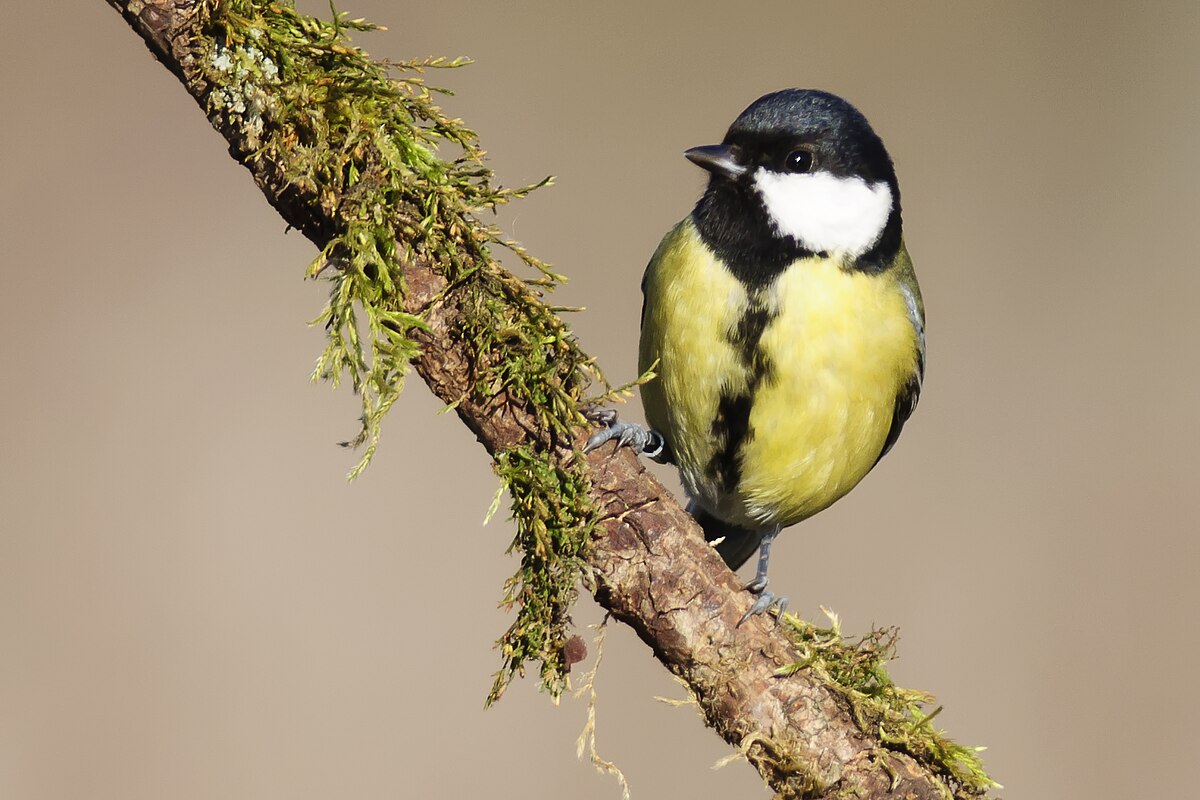}
\includegraphics[width=2.9in]{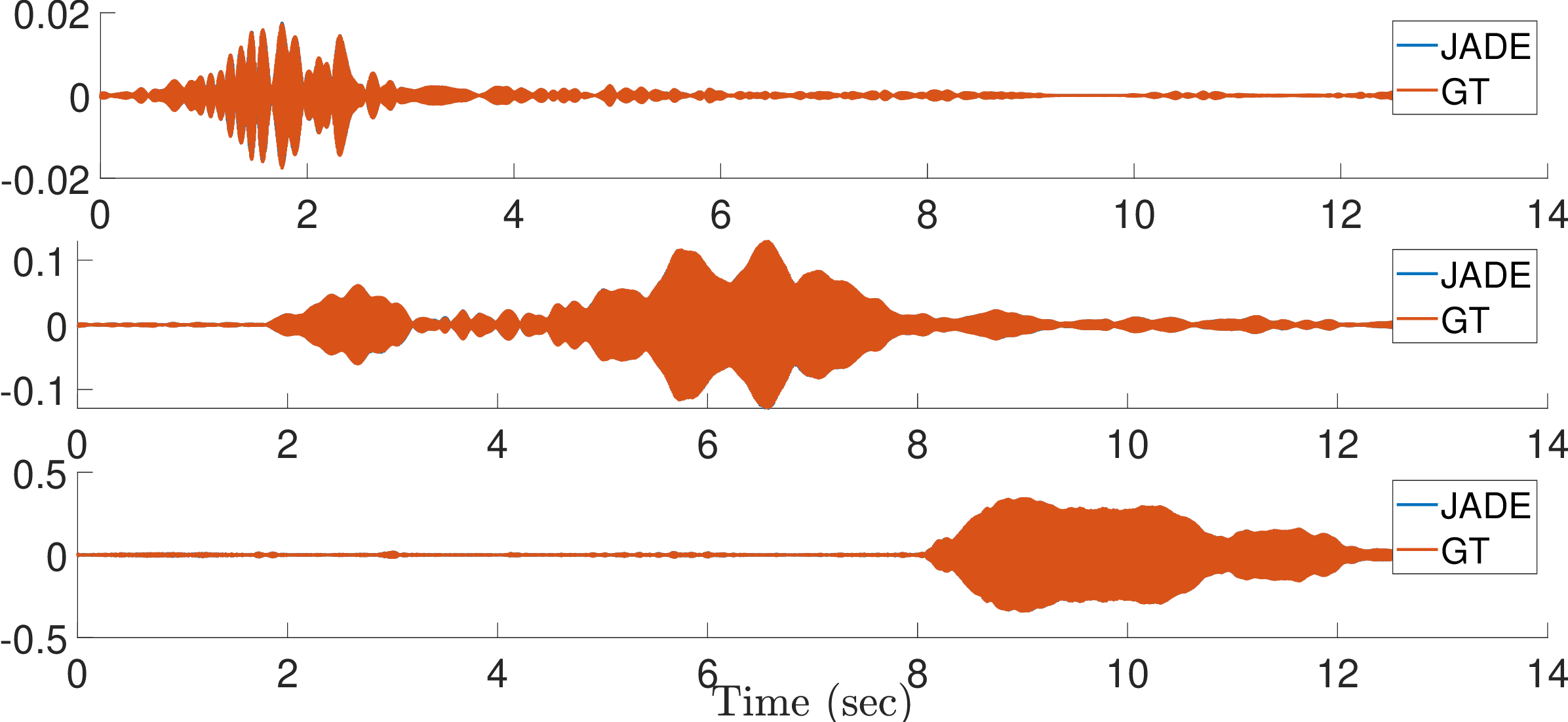}
\caption{Example 5. In the left panel, a picture of a parus major also known as cinciallegra (Italian) or great tit (English). (2024, April 14). In Wikipedia. In the right panel, the subplots show the first three IMFs (red) obtained from the FIF decomposition of the full bird chirp signal. The reconstruction obtained from the JADE phase estimate is shown in blue.}
\label{birdFullReconstr}
\end{figure}

\begin{figure}[ht] 
\centering
\includegraphics[width=0.8\textwidth]{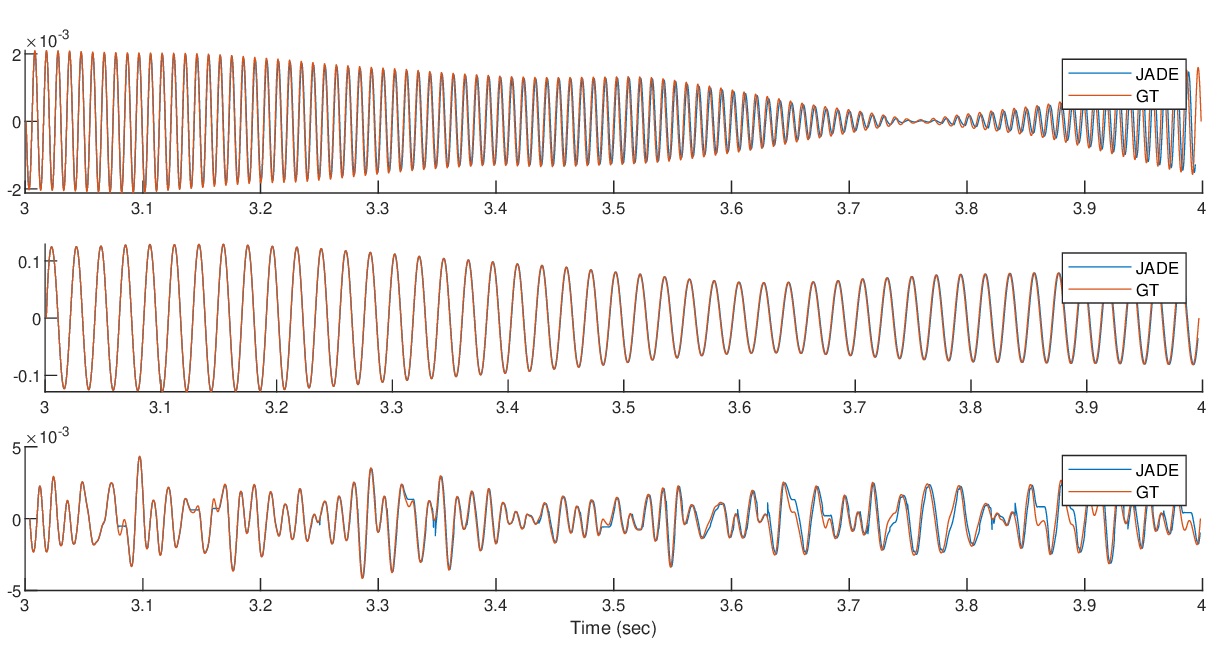}
\caption{Example 5. The subplots show a zoomed-in version of the first three IMFs (red) obtained from the FIF decomposition of a portion of the bird chirp signal. The reconstruction obtained from the JADE phase estimate is shown in blue.}
\label{birdPortionReconstr}
\end{figure}

We consider as a first example the chirping sound produced by the bird parus major also known as cinciallegra or great tit, ref. left panel of Fig.~\ref{birdFullReconstr} retrieved from Wikipedia\footnote{\url{https://it.wikipedia.org/wiki/Parus_major}}. We consider a 14-second recording taken from an online repository\footnote{The audio recording was downloaded from the website \url{https://pixabay.com/it/sound-effects/search/cinciallegra/}}, and we decompose it via FIF. The first three IMFs obtained are shown in the right panel of Fig.~\ref{birdFullReconstr}. On top of them, we show the components reconstructed using the phase computed via JADE for each IMF. In Fig.~\ref{birdPortionReconstr} we show a zoomed-in version of 1 second. From this second figure, we can see that JADE allows us to reconstruct with good accuracy the phase, and consequently, the components, contained in the original signal.

\clearpage

\subsection{Example 6: Gravitational Wave Analysis}

\begin{figure}[ht] 
\centering
\includegraphics[width=0.8\textwidth]{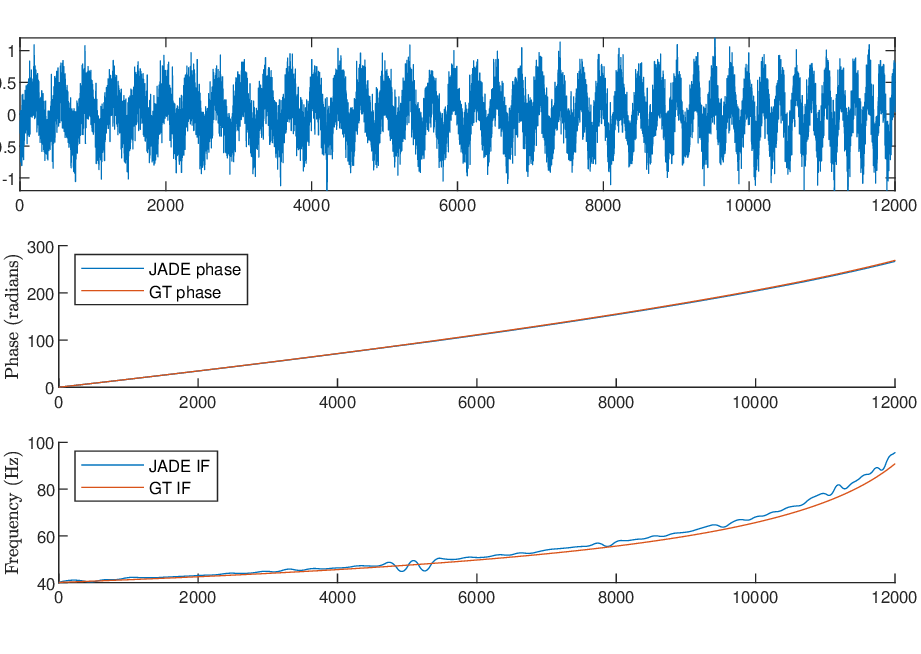}
\caption{Example 6. Simulated inspiral GW signal with noise (top). JADE phase estimate versus ground truth phase (middle). JADE IF estimate versus ground truth IF (bottom).}
\label{fig:gravWave}
\end{figure}

In this example, we consider the phase estimation of a simulated inspiral gravitational wave (GW) signal. Inspiral GW signals are generated during the end-of-life merging of neutron star or black hole binary systems. There exist many approximate analytical models to simulate inspiral GW waveforms; we use \textit{SEOBNRv4} \cite{Bohe2017}, an effective-one-body (EOB) model employed by the LIGO-VIRGO collaboration. Using the PyCBC software package \cite{pycbc}, we simulate an inspiral GW signal using \textit{SEOBNRv4}, with a ringdown time step of 1/8192 and a lower frequency of 40Hz. This simulated waveform is shown in the upper panel of Fig.~\ref{fig:gravWave}. The ground truth phase and IF are obtained directly from PyCBC. The middle panel shows the JADE phase estimate versus the ground truth phase, and the bottom panel shows the JADE IF estimate versus the ground truth IF. In this last panel we observe small oscillations appearing. This is an effect due to the application of the derivative operator, which is known in mathematics to amplify the noise present in a curve.

\begin{figure}[ht] 
\centering
\includegraphics[width=0.8\textwidth]{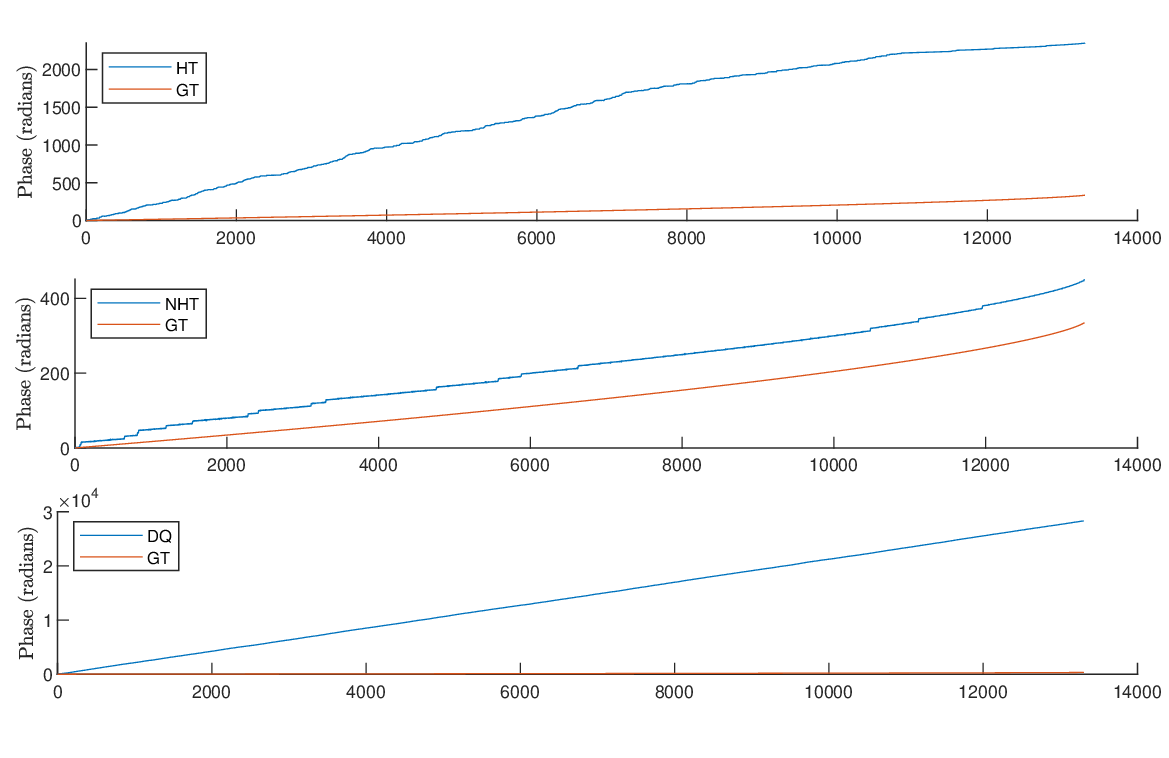}
\caption{Example 6. Comparison of phase estimates (in blue) from Hilbert Transform (top), Normalized Hilbert Transform (middle), and Direct Quadrature (bottom) versus the ground truth (in red) for the gravitational wave signal.}
\label{fig:gravWaveComparison}
\end{figure}

In Fig.~\ref{fig:gravWaveComparison}, we show the comparison of phase estimates from HT, NHT, and DQ methods on the gravitational wave signal. As seen previously, these methods fail to produce the correct phase estimate in noisy conditions.

Because the ground truth phase and IF are available from the PyCBC module, we test the JADE estimate performance in different noise conditions. We add Gaussian noise to the original waveform to create a range of SNR signals and measure the relative error between the JADE phase estimate and ground truth phase, defined in (\ref{eq:error}), for each case. The results are summarized in Table \ref{table_err_SNR_grav}.

\begin{table}[ht!]
\begin{center}
\begin{tabular}{||c c c c c c c||}
 \multicolumn{7}{|c|}{\textbf{SNR (dB)}} \\
 \hline
 &6.02 & 3.63 & 1.59 & -1.31 & -3.41 & -6.58 \\ [0.5ex]
 \hline\hline
 \textbf{$\epsilon$}&7.2e-4 & 5.5e-4 & 1.1e-3 & 1.4e-3 & 1.5e-3 & 1.8e-3\\
 \hline
\end{tabular}
\end{center}
\caption{Example 6. Relative error in phase estimate (\textbf{$\epsilon$}) versus noise level for the simulated gravitational waveform.}
\label{table_err_SNR_grav}
\end{table}

The -6.58 dB case is shown in Fig.~\ref{fig:gravWave}. We analyze the latter portion of the signal, which more clearly shows the effect of the noise on the gravitational waveform.

\subsection{Example 7: Electrocardiogram Analysis}

\begin{figure}[ht] 
\centering
\includegraphics[width=0.8\textwidth]{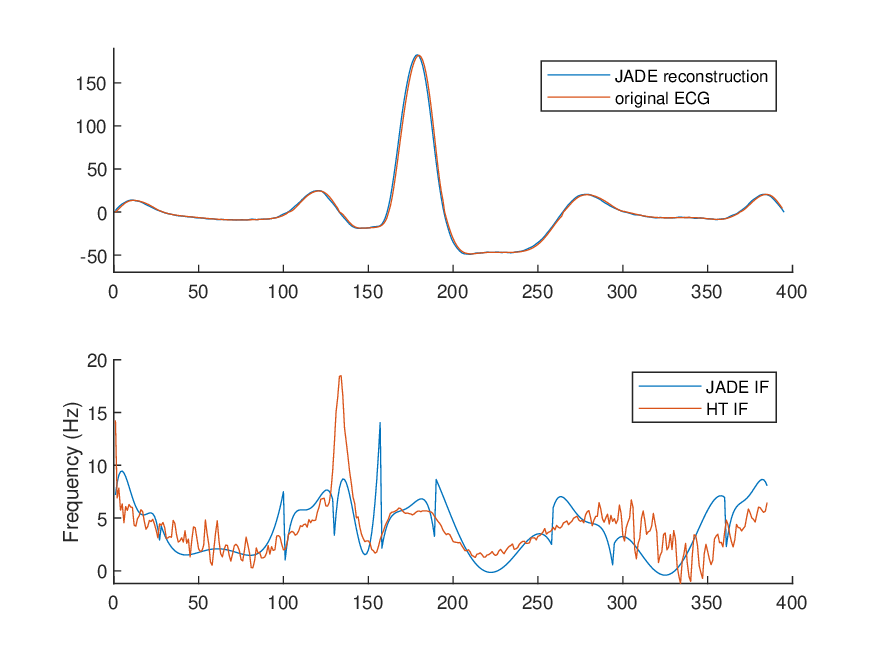}
\caption{Example 7. Top panel shows ECG pulse from MIT-BIH Arrhythmia database (red) and reconstruction of the signal from JADE phase estimate (blue). The bottom panel shows the JADE IF estimate (blue) and HT IF (red).}
\label{ECGreconstr}
\end{figure}

We next consider the phase estimation and reconstruction of electrocardiogram (ECG) signals from the MIT-BIH Arrhythmia database publicly available at \textit{PhysioNet} \cite{ecgdb}. This database consists of 48 two-channel ECG recordings of approximately 30 minutes in length, from 47 different patients. The recordings were sampled at 360Hz. From a particular recording (serial number 105), we isolate a singular ECG pulse of a duration of 0.7 seconds. We apply a moving average filter in order to remove the high-frequency component from the signal. The resulting signal is shown in the top panel of Fig.~\ref{ECGreconstr} in red. The reconstruction $\widehat{ECG}$ computed from Equation \eqref{eq:reconstruction} is shown in the top panel of Fig.~\ref{ECGreconstr} in blue. It can be seen that the reconstruction based on the JADE phase estimate accurately represents the original ECG signal. In the bottom panel, we show the JADE IF estimate along with the HT IF for comparison. It is clear from this last plot that there is still some noise left in the phase curve reconstructed via JADE. The derivative is an operator that amplifies the noise. In this case, the IF obtained via derivation of the JADE phase and HT are comparable.

\subsection{Example 8: Magnetic Field Response Analysis}

\begin{figure}[ht] 
\centering
\includegraphics[width=0.8\textwidth]{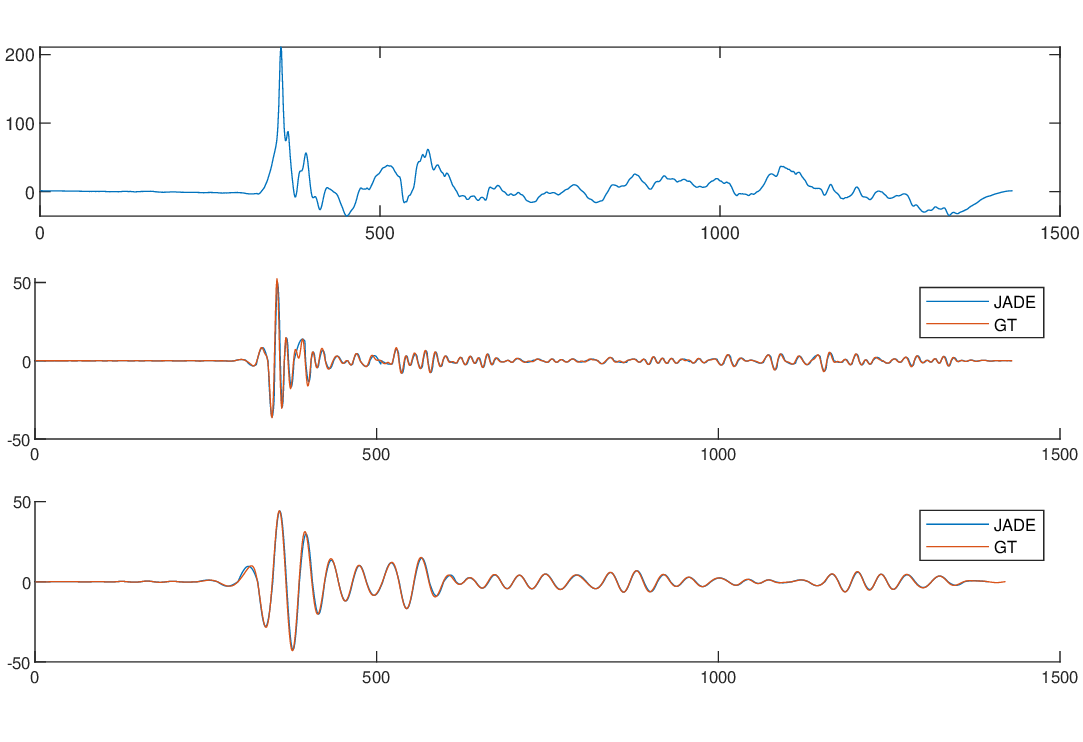}
\caption{Example 8. Top panel shows the original geomagnetic response waveform. The middle and bottom panels show the second and third IMFs respectively (red), and the corresponding reconstructions using JADE phase estimates (blue).}
\label{Magreconstr}
\end{figure}

In this last example, we consider phase estimation of magnetospheric and geomagnetic response to solar wind. Collision of the solar wind with the magnetic field causes a magnetospheric response characterized by sudden enhancement of the magnetic field intensity \cite{piersantigeophysical}. We analyze one particular event detected by GEOS (Geostationary Operational Environmental Satellites) spacecraft; this signal is shown in the upper panel of Fig.~\ref{Magreconstr}. First, the signal is decomposed into IMFs using the FIF method. We estimate the phase of the second and third IMFs using JADE and compute the signal reconstruction according to (\ref{eq:reconstruction}) for both IMFs. The results are shown in the middle and bottom panels of Fig.~\ref{Magreconstr}.

\clearpage

\subsection{Example 9: acoustic emission of mechanical engine}

\begin{figure}[ht] 
\centering
\includegraphics[width=0.8\textwidth]{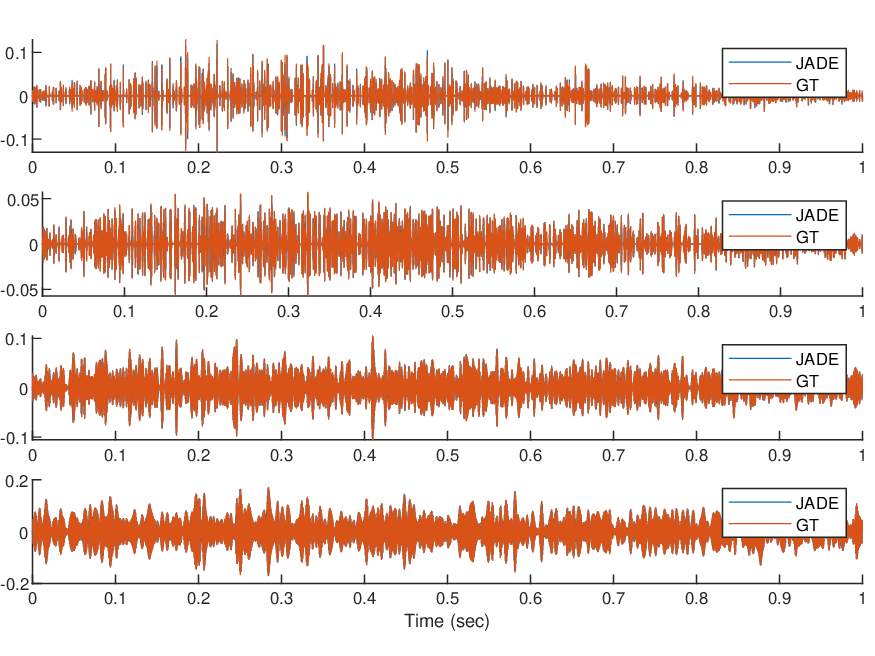}
\caption{Example 9. The subplots show four IMFs (red) obtained from the FIF decomposition of 1 second of acoustic emission of a revving engine. The reconstruction obtained from the JADE phase estimate is shown in blue.}
\label{fig:Engine_IMFs}
\end{figure}

\begin{figure}[ht] 
\centering
\includegraphics[width=0.8\textwidth]{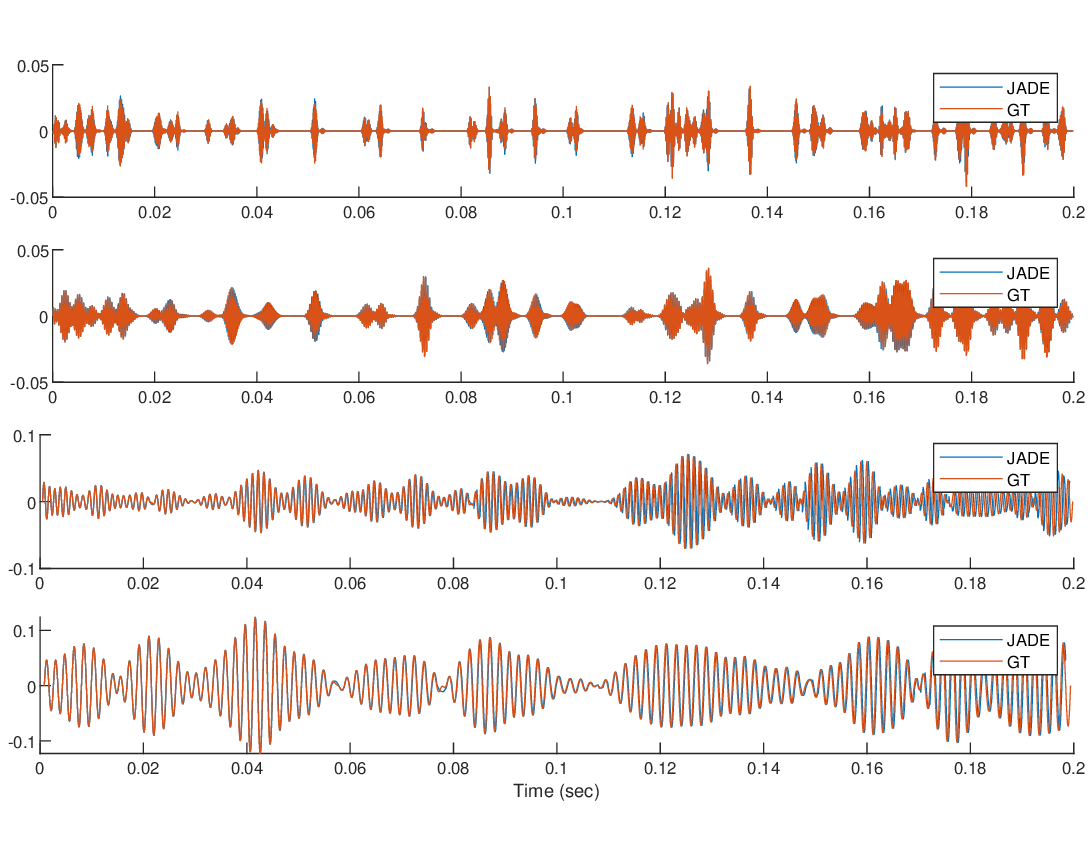}
\caption{Example 9. The first 0.2 seconds of the previous plots are shown.}
\label{fig:Engine_IMFs_Portion}
\end{figure}

The acoustic emissions of moving parts of mechanical systems can be used in the diagnostic, classification,
and identification of these systems. The topic is of interest as confirmed by recent papers, see for
example \cite{czarnecki2016instantaneous}. In this example, we consider the acoustic emission of the engine of a Lamborghini Gallardo\footnote{A sample of recording was downloaded from the on line database available at \url{https://freesound.org/s/222607/} (date of download:
December,2024).}. The sampling rate of the track is equal to 44100 samples per second in a lossless waveform audio file format (WAVE or WAV) and the precision is 16 bits per sample. We consider one second of the recording of the acoustic emission of the Lamborghini Gallardo engine when its speed is increased as the accelerator is pressed, the so-called rev phase. In Fig. \ref{fig:Engine_IMFs} and FIg. \ref{fig:Engine_IMFs_Portion}, a sample of the decomposition via FIF is presented together with the reconstructed curves obtained via JADE. The analyzed signal is multicomponent and nonstationary.

\begin{figure}[ht] 
\centering
\includegraphics[width=0.8\textwidth]{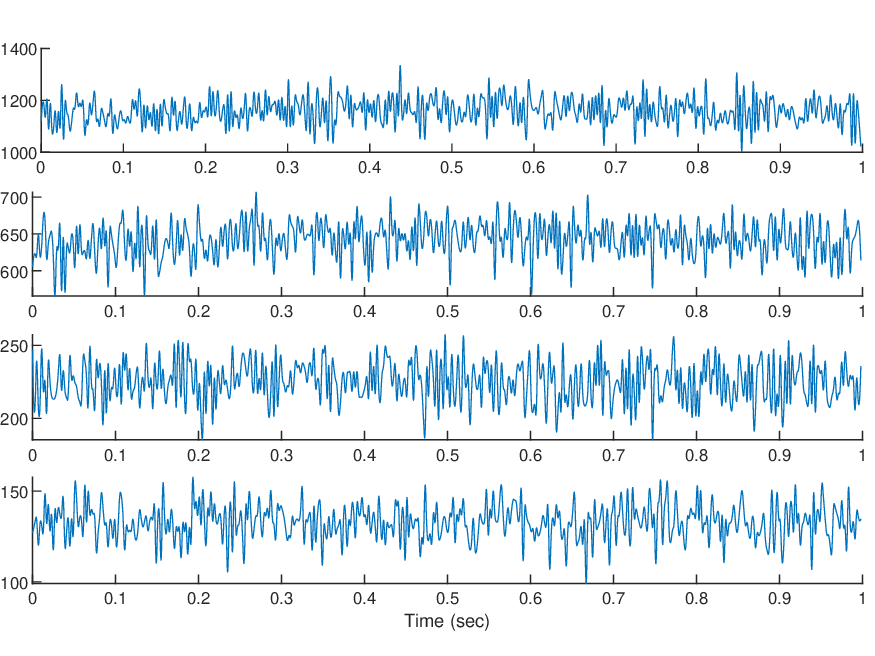}
\caption{Example 9. The IF curves obtained from JADE of 1 second of acoustic emission of a revving engine.}
\label{fig:Engine_IFs}
\end{figure}

In Fig. \ref{fig:Engine_IFs} we report the IF curves computed via JADE and associated with the 4 IMFs shown in Fig. \ref{fig:Engine_IMFs}. Also in this case, the derivative amplifies the noise present in the reconstructed phase curves.

\section{Conclusion}

Many real-life applications require the analysis of non-stationary signals whose frequencies vary rapidly over time. In recent years, many innovative and nonlinear approaches have been proposed for the decomposition of such signals into mono-component signals, also known in the field as IMFs. In this article, we briefly review and apply the Fast Iterative Filtering (FIF) technique. Once the signals have been decomposed into IMFs, we need to study their phase and frequency content over time with, possibly, high accuracy.
In this work, we propose a new approach, called JADE, based on the Dynamic Time Warping method, for the estimation of the instantaneous phase and frequency of a mono-component signal. We test this method on both synthetic and real-life signals, comparing results with other algorithms proposed so far in the literature. The results show clearly that JADE outperforms any other method developed so far in the literature and proves to be stable, even in the presence of heavy noise. Regarding the instantaneous frequency estimation, we observed from the tests conducted that the derivative operator tends to amplify small oscillations that are naturally present in the IMFs produced by FIF and, as a consequence, in the instantaneous phase curves produced via the proposed technique. We plan to study in a future work how to reduce this effect.

\section{Acknowledgments}
A. Cicone is member of the Gruppo Nazionale Calcolo Scientifico-Istituto Nazionale di Alta Matematica (GNCS-INdAM), was partially supported through the GNCS-INdAM Project CUP E53C23001670001, and was supported by the Italian Ministry of the University and Research and the European Union through the ``Next Generation EU'', Mission 4, Component 1, under the PRIN PNRR 2022 grant number CUP E53D23018040001 ERC field PE1 project P2022XME5P titled ``Circular Economy from the Mathematics for Signal Processing prospective''.

\clearpage

\bibliographystyle{elsarticle-harv}

\end{document}